\documentclass[11pt,a4paper]{article}
\pdfoutput=1
\usepackage{jheppub}
\usepackage{cite}

\usepackage{multirow, graphicx,amssymb,url,mathrsfs,amsmath}
\usepackage{wrapfig,boxedminipage,subfigure,epsfig}
\usepackage{amsxtra,amstext,latexsym,dsfont,amsfonts}
\usepackage{color}
\usepackage[dvipsnames]{xcolor}
\usepackage{float}
\usepackage{slashed}
\usepackage{calligra}
\DeclareFontShape{T1}{calligra}{m}{n}{<->s*[2.2]callig15}{}
\DeclareMathAlphabet{\mathcalligra}{T1}{calligra}{m}{n}

\newcommand{\be}{\begin{equation}}
\newcommand{\ee}{\end{equation}}
\newcommand{\bea}{\begin{eqnarray}}
\newcommand{\eea}{\end{eqnarray}}

\newcommand{\pa}{\partial}

\title{Regge conformal blocks from the Rindler-AdS black hole and 
the pole-skipping phenomena}

\author[]{Keun-Young Kim,}
\author[]{Kyung-Sun Lee,}
\author[]{and Mitsuhiro Nishida}

\affiliation[]{School of Physics and Chemistry, Gwangju Institute of Science and Technology, 123 Cheomdan-gwagiro, Gwangju 61005, Korea}

\emailAdd{fortoe@gist.ac.kr}
\emailAdd{kyungsun.cogito.lee@gmail.com}
\emailAdd{mnishida@gist.ac.kr}

\abstract{We study a holographic construction of conformal blocks in the Regge limit of four-point scalar correlation functions by using coordinates of the two-sided Rindler-AdS black hole. As a generalization of geodesic Witten diagrams, we construct diagrams with four external scalar fields in the Rindler-AdS black hole by integrating over two half-geodesics between the centers of Penrose diagrams and points at the AdS boundary. We demonstrate that late-time behaviors of the diagrams coincide with the Regge behaviors of conformal blocks. We also point out their relevance with  
the pole-skipping phenomena by showing that the near-horizon analysis of symmetric traceless fields with any integer spin in the Rindler-AdS black hole can capture the Regge behaviors of conformal blocks.
 }

\begin{document}
\maketitle



\section{Introduction}
The singularity structure of Euclidean correlators in quantum field theories is determined by the operator product expansion (OPE), which is an important concept in the local theories. On the other hand, we often encounter the singular behaviors in the theories with the Lorentzian signature that are different from the ones in the Euclidean theories. One example is the Regge limit of four-point functions in conformal field theories (CFTs) \cite{Brower:2006ea, Cornalba:2006xk, Cornalba:2006xm, Cornalba:2007zb, Cornalba:2007fs, Cornalba:2008qf, Costa:2012cb}, which describes scattering at high energies. The behaviors of correlation functions in the Regge limit are well constrained by unitarity, analyticity, and causality (see, for example, \cite{Camanho:2014apa, Afkhami-Jeddi:2016ntf, Caron-Huot:2017vep,  Kulaxizi:2017ixa, Costa:2017twz}).

An application of the Regge limit is to calculate out-of-time-order correlation functions (OTOCs) \cite{larkin1969quasiclassical, Kitaev-2014}. If Euclidean correlation functions in CFTs are given, one can compute the OTOCs by using an analytic continuation of the Euclidean correlators to the Regge limit \cite{Roberts:2014ifa, Perlmutter:2016pkf}. The Lyapunov exponent of the OTOCs with four scalar fields in large $N$ theories is a diagnosis of quantum chaos, and the consistency of the theories in the Regge limit bounds the Lyapunov exponent \cite{Maldacena:2015waa}\footnote{See also \cite{Chandorkar:2021viw} for the constraint in the bulk flat space S matrix from the chaos bound.}.

Recently, it has been proposed that the Lyapunov exponent and the butterfly velocity in maximally chaotic systems can be deduced from a ``pole-skipping" point of the retarded Green's function of the energy-momentum tensor \cite{Grozdanov:2017ajz, Blake:2017ris}. Here, the pole-skipping points are points such that Green's functions in the momentum space are not uniquely determined. This connection between the pole structure and quantum chaos is called the ``pole-skipping phenomenon". A holographic method to find the pole-skipping points from equations of motion on the horizon of black holes, which is called ``near-horizon analysis", has been developed \cite{Blake:2018leo}.

When the energy-momentum tensor exchange is dominant in the OTOCs \cite{Perlmutter:2016pkf, Afkhami-Jeddi:2017rmx}, the Lyapunov exponent and the butterfly velocity of the holographic CFTs in the Rindler space can be computed from the exponential behavior of the Regge conformal block with the energy-momentum tensor exchange, which is  defined by the Regge  limit of the conformal block.  The pole-skipping points of the energy density operator in the Rindler space, which are derived by a CFT method \cite{Haehl:2019eae} and by the near-horizon analysis in the  Rindler-AdS black hole \cite{Ahn:2019rnq}, are consistent with the exponential behavior of the Regge conformal block, which means the pole-skipping phenomenon of the energy-momentum tensor in the Rindler space. The pole-skipping points of scalar and vector fields\footnote{In two-dimensional CFTs, the pole-skipping points of a spin-3 conserved current \cite{Haehl:2018izb} and higher-spin conserved currents \cite{Das:2019tga} has been observed.} in the Rindler space can also capture the exponential behaviors of the Regge conformal blocks with the scalar and vector exchange \cite{Ahn:2020bks}.

To understand why the near-horizon analysis in the Rindler-AdS black hole can provide the Regge behaviors of conformal blocks, it would be useful to represent the Regge conformal blocks in the bulk picture, and holographic representations of the Regge conformal or OPE blocks have been studied in \cite{Cornalba:2006xk, Cornalba:2006xm, Cornalba:2007zb, Cornalba:2007fs, Afkhami-Jeddi:2017rmx, Kobayashi:2020kgb}. For a clear interpretation of the near-horizon analysis, we want to construct the Regge conformal blocks by using Kruskal coordinates of the Rindler-AdS black hole explicitly such as the shock wave computations in \cite{Shenker:2013pqa, Roberts:2014isa, Shenker:2014cwa}.

In this paper, we construct scattering diagrams with four external scalar fields in the two-sided Rindler-AdS black hole for the Regge conformal blocks, which are generalizations of geodesic Witten diagrams in Euclidean AdS$_{d+1}$ for Euclidean conformal blocks \cite{Hijano:2015zsa}. In our construction, instead of two complete geodesics between boundary points in the usual geodesic Witten diagrams, we integrate over two half-geodesics between the boundary points and the centers of Penrose diagrams. We call our diagrams ``half-geodesic Witten diagrams"\footnote{Note that our notion of the half-geodesic Witten diagrams is different from the ones in \cite{Kulaxizi:2018dxo, Chen:2019fvi, David:2020fea} that are integrated over one complete geodesic.} and show that their late-time behaviors agree with the Regge behaviors of conformal blocks.

We also show that the near-horizon analysis of symmetric traceless fields with spin $\ell$ in the Rindler-AdS black hole can obtain equations for the Regge conformal blocks and discuss a connection between the near-horizon analysis and integrals of bulk-to-bulk propagators in the half-geodesic Witten diagrams. 
Our analysis is an extension of the near-horizon analysis in \cite{Ahn:2020bks} to arbitrary integer spin and to a finite spatial distance.

The paper is organized as follows. We review the Euclidean conformal blocks and the geodesic Witten diagrams and compare their asymptotic behaviors in Section \ref{sec2}. In Section \ref{sec3}, we construct the half-geodesic Witten diagrams and compute their late-time behaviors. We perform the near-horizon analysis of spin-$\ell$ fields in Section \ref{Reggeeomnha}.
Section \ref{sec5} concludes the paper and discusses future work.

\section{Asymptotic behaviors of the Euclidean conformal blocks via geodesics}\label{sec2}
As a warm-up, we study the Euclidean conformal blocks by evaluating the geodesic Witten diagrams in Euclidean $\text{AdS}_{d+1}$, which are the gravity duals of the Euclidean conformal blocks with power-law prefactors \cite{Hijano:2015zsa}. In particular, the geodesic Witten diagrams and the conformal blocks with the prefactors have the same asymptotic behaviors in small cross ratios \cite{Hijano:2015zsa, Dyer:2017zef}. From this property, we see that the asymptotic behaviors of conformal blocks can be estimated from classical equations of exchange fields on a geodesic without using exact solutions of bulk-to-bulk propagators.

\subsection{Review of the Euclidean conformal blocks and the geodesic Witten diagrams}
We start reviewing the Euclidean conformal blocks and the geodesic Witten diagrams \cite{Hijano:2015zsa}. Consider a CFT four-point function $\langle \mathcal{O}_1(x_1)\mathcal{O}_2(x_2)\mathcal{O}_3(x_3)\mathcal{O}_4(x_4)\rangle$ of scalar primary operators $\mathcal{O}_a(x_a)$ with conformal dimension $\Delta_a$ in $d$-dimensional Euclidean space $\mathbb{R}^{d}$. Thanks to the conformal symmetry, the four-point function can be expanded in terms of conformal blocks $G_{\Delta, \ell}(z, \bar{z})$:
\begin{align}
&\langle \mathcal{O}_1(x_1)\mathcal{O}_2(x_2)\mathcal{O}_3(x_3)\mathcal{O}_4(x_4)\rangle=\sum_\mathcal{O}C_{12\mathcal{O}}C_{34\mathcal{O}}W_{\Delta, \ell}(x_a)\,,\\
&W_{\Delta, \ell}(x_a):=\left(\frac{x_{24}^2}{x_{14}^2}\right)^{\frac{1}{2}(\Delta_1-\Delta_2)}\left(\frac{x_{14}^2}{x_{13}^2}\right)^{\frac{1}{2}(\Delta_3-\Delta_4)}\frac{G_{\Delta, \ell}(z, \bar{z})}{(x_{12}^2)^{\frac{1}{2}(\Delta_1+\Delta_2)}(x_{34}^2)^{\frac{1}{2}(\Delta_3+\Delta_4)}}\, ,\label{cbwp}
\end{align}
where $x_{ab}:=x_a-x_b$, $C_{ab\mathcal{O}}$ is the OPE coefficient, $\Delta$ and $\ell$ are the conformal dimension and spin of primary field $\mathcal{O}$, and cross ratios $z$ and $\bar{z}$ obey
\begin{align}
z\bar{z}=\frac{x_{12}^2x_{34}^2}{x_{13}^2x_{24}^2}\,, \;\;\; (1-z)(1-\bar{z})=\frac{x_{14}^2x_{23}^2}{x_{13}^2x_{24}^2}\, .\label{cr}
\end{align}
The asymptotic behaviors of $G_{\Delta, \ell}(z, \bar{z})$ in small cross ratios $z, \bar{z}\to 0$ are \cite{Dolan:2003hv, Costa:2011dw}
\begin{align}
G_{\Delta, \ell}(z, \bar{z})\propto (z\bar{z})^{\frac{\Delta}{2}}C_{\ell}^{d/2-1}\left(\frac{z+\bar{z}}{2\sqrt{z\bar{z}}}\right) \;\;\; (z, \bar{z}\to 0)\, , \label{abec}
\end{align}
where $C_{\ell}^{d/2-1}(x)$ is the Gegenbauer polynomial.

\begin{figure}
\centering
     {\includegraphics[width=4cm]{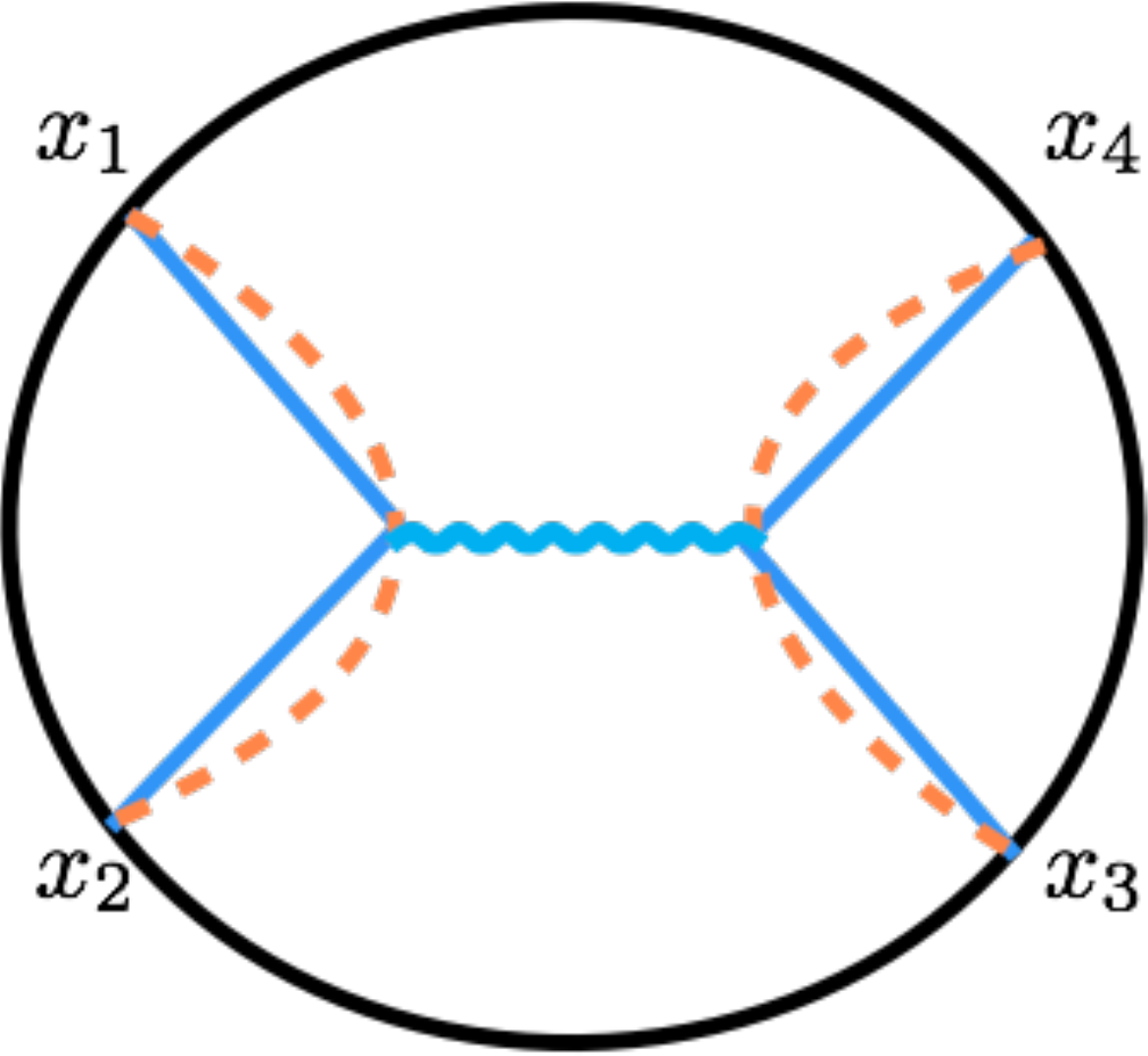}}
 \caption{Geodesic Witten diagram in Euclidean AdS$_{d+1}$. A circle represents the AdS boundary, four straight lines are the scalar bulk-to-boundary propagators, and a wavy line is a bulk-to-bulk propagator. Two dashed curves are the geodesics between the boundary points, and the integrals in (\ref{gwd}) are over the two geodesics.   }\label{gwdeads}
\end{figure}

The geodesic Witten diagrams, which are diagrams in Euclidean $\text{AdS}_{d+1}$ with integration of vertices over two geodesics between boundary points as shown in Fig.~\ref{gwdeads}, give integral representations of $W_{\Delta, \ell}(x_a)$ (\ref{cbwp}) via AdS propagators . The scattering amplitudes of the geodesic Witten diagrams $\mathcal W_{\Delta,\ell}$ with four external scalar fields are given by
\begin{align}
\mathcal W_{\Delta,\ell}:=\int_{\gamma_{12}} d\lambda \int_{\gamma_{34}} d\lambda' &G_{b\partial}\left(y(\lambda), x_1; \Delta_1\right)G_{b\partial}\left(y(\lambda), x_2; \Delta_2\right)G_{bb}\left(y(\lambda) , y(\lambda'); \Delta, \ell\right)\notag\\
\times&G_{b\partial}\left(y(\lambda'), x_3; \Delta_3\right)G_{b\partial}\left(y(\lambda'), x_4; \Delta_4\right)\, ,\label{gwd}
\end{align} 
where $\gamma_{ab}$ are the geodesics between the boundary points $x_a$ and $x_b$, and $\lambda$ and $\lambda'$ are proper length parameters of $\gamma_{12}$ and $\gamma_{34}$, respectively.  In the construction, we use the scalar bulk-to-boundary propagators $G_{b\partial}\left(y, x_a; \Delta_a\right)$ and the pulled-back bulk-to-bulk propagator $G_{bb}\left(y(\lambda), y(\lambda'); \Delta, \ell\right)$ of a spin-$\ell$ symmetric traceless field.  The authors of \cite{Hijano:2015zsa} showed that $\mathcal W_{\Delta,\ell}$ (\ref{gwd}) coincides with $W_{\Delta, \ell}(x_a)$ (\ref{cbwp}) up to normalization. In particular, $\mathcal W_{\Delta,\ell}$ (\ref{gwd}) satisfies the properties that $W_{\Delta, \ell}(x_a)$ must satisfy.

The explicit forms of geodesics and propagators are given as follows. We use Poincar\'e coordinates $y=\{\eta, x\}=\{\eta, w, \bar{w}, \mathbf{x}_\perp\}$ with a metric
\begin{align}
ds^2=\frac{d\eta^2+dx^2}{\eta^2}=\frac{d\eta^2+dwd\bar{w}+d\mathbf{x}_\perp^2}{\eta^2}\, ,\label{poincare}
\end{align}
where the AdS radius is set to one.
The geodesic between the boundary points $x_a$ and $x_b$ in the Poincar\'e coordinates is a semi-circle
\begin{align}
\eta^2+\left(x-\frac{x_a+x_b}{2}\right)^2=\left(\frac{x_{ab}}{2}\right)^2\ .
\end{align}
This geodesic is parametrized by the proper parameter $\lambda$ \cite{Hijano:2015zsa}
\begin{align}
x^i(\lambda)&=\frac{x_a^i+x_b^i}{2}-\frac{x_{ab}^i}{2}\tanh\lambda, \;\;\; \eta(\lambda)=\frac{(x_{ab}^2)^{\frac{1}{2}}}{2\cosh\lambda}\, .\label{geo}
\end{align} 
If $x_a^i=x_b^i$, $x^i(\lambda)$ is a constant on the geodesic. The AdS scalar propagators are given by \cite{Costa:2014kfa}
\begin{align}
&G_{b\partial}\left(y, x_a; \Delta_a\right)=\mathcal{C}_{\Delta_a, 0}\left(\frac{\eta}{\eta^2+(x-x_a)^2}\right)^{\Delta_a}\, ,\label{sbubap}\\
&G_{bb}\left(y, y'; \Delta, 0\right)=\mathcal{C}_{\Delta, 0}\left(\frac{\xi}{2}\right)^\Delta\,_2F_1\left(\frac{\Delta}{2}\, , \frac{\Delta+1}{2}, \Delta+1-\frac{d}{2}; \xi^2\right),\label{sbubup}\\
&\xi:=\frac{2\eta\eta'}{\eta^2+\eta'^2+(x-x')^2}\, , \;\;\; \mathcal{C}_{\Delta, \ell}:=\frac{(\ell+\Delta-1)\Gamma(\Delta)}{2\pi^{d/2}(\Delta-1)\Gamma(\Delta+1-d/2)}\, .
\end{align}
The bulk-to-bulk propagators with nonzero $\ell$ are constructed in \cite{Costa:2014kfa} using the embedding formalism.

\subsection{Asymptotic behavior of the scalar exchange geodesic Witten diagram} \label{sec22}
Let us rederive the asymptotic behavior (\ref{abec}) with $\ell=0$ from the geodesic Witten diagram representation (\ref{gwd}).
For simplicity, we use conformal transformations to set
\begin{align}
x_1=(\infty, \infty, \mathbf{0}_\perp)\,, \ \
x_2=(0, 0, \mathbf{0}_\perp)\,, \ \
x_3=(1-z, 1-\bar{z}, \mathbf{0}_\perp)\,, \ \
x_4=(1, 1, \mathbf{0}_\perp)\,.\label{sp}
\end{align}
One can check that (\ref{sp}) satisfies the conditions (\ref{cr}).  To derive the asymptotic behavior of $\mathcal W_{\Delta, 0}$, it is enough to compute the $z\bar{z}$-dependent part
\begin{align}
G_{bb}\left(y(\lambda), y(\lambda'); \Delta, 0\right)G_{b\partial}\left(y(\lambda'), x_3; \Delta_3\right)G_{b\partial}\left(y(\lambda'), x_4; \Delta_4\right)\, ,
\end{align}
since the other two $G_{b\partial}$s are not functions of $z$ and $\bar{z}$ in the coordinates \eqref{sp}.
We first focus on the bulk-to-boundary propagators $G_{b\partial}\left(y(\lambda'), x_3; \Delta_3\right)G_{b\partial}\left(y(\lambda'), x_4; \Delta_4\right)$.
By using (\ref{geo}) and (\ref{sbubap}), we obtain
\begin{equation}
\begin{split}
&G_{b\partial}\left(y(\lambda'), x_3; \Delta_3\right)G_{b\partial}\left(y(\lambda'), x_4; \Delta_4\right)\\
\,=\,&\mathcal{C}_{\Delta_3, 0}\mathcal{C}_{\Delta_4, 0}\frac{e^{-(\Delta_3-\Delta_4)\lambda'}}{(z\bar{z})^{\frac{1}{2}(\Delta_3+\Delta_4)}}\propto\frac{1}{(z\bar{z})^{\frac{1}{2}(\Delta_3+\Delta_4)}}\, .\label{B2b result}
\end{split}
\end{equation}
This $z\bar{z}$-dependence corresponds to the prefactor $1/(x_{34}^2)^{\frac{1}{2}(\Delta_3+\Delta_4)}$ in (\ref{cbwp}). The other prefactors in (\ref{cbwp}) are numbers independent of $z$ and $\bar{z}$.

We next evaluate the bulk-to-bulk propagator $G_{bb}\left(y(\lambda), y(\lambda'); \Delta, 0\right)$. In the limit of small cross ratios, $\xi$ between $y(\lambda)$ and $y(\lambda')$ becomes small as
\begin{equation}
\begin{split}
y(\lambda')=(\eta(\lambda'), x(\lambda')) \to \left(\frac{(z\bar{z})^{\frac{1}{2}}}{2\cosh\lambda'}, x_4\right), & \\
\xi\to\frac{2\eta(\lambda)\eta(\lambda')}{\eta(\lambda)^2+(x(\lambda)-x_4)^2} = f(\lambda, \lambda')(z\bar{z})^{\frac{1}{2}} \propto (z\bar{z})^{\frac{1}{2}}& \;\;\; (z,\bar{z}\to0)\,, \label{giscl}
\end{split}
\end{equation}
where we use $y(\lambda)\ne y(\lambda')$, and $f(\lambda, \lambda')$ is some function, which will be integrated at the end. Therefore, the asymptotic behavior of $G_{bb}\left(y(\lambda), y(\lambda'); \Delta, 0\right)$ is
\begin{align}
G_{bb}\left(y(\lambda), y(\lambda'); \Delta, 0\right)\simeq \mathcal{C}_{\Delta, 0}\left(\frac{\xi}{2}\right)^\Delta\propto (z\bar{z})^{\frac{1}{2}\Delta} \;\;\; (z,\bar{z}\to0)\,.\label{absp}
\end{align}
This behavior matches with the asymptotic behavior of conformal block (\ref{abec}) with $\ell=0$. Note that the integrations over $\lambda$ and $\lambda'$ are factored out in the limit $z,\bar{z}\to0 $ so do not affect the $z\bar{z}$-dependence in the asymptotic behavior of $\mathcal W_{\Delta, 0}$.

\subsection{Asymptotic behaviors of the spin-$\ell$ exchange geodesic Witten diagrams:  index-free polynomials}\label{subsec:gwdsl}

We can also study the asymptotic behaviors of the geodesic Witten diagrams with nonzero $\ell$. For a systematic analysis, we introduce embedding formalism \cite{Costa:2014kfa}. Euclidean $\text{AdS}_{d+1}$ and Euclidean space $\mathbb{R}^{d}$ can be embedded into Minkowski space $\mathbb{R}^{1, d+1}$ with a  metric
\begin{align}
ds^2=-dY^+dY^-+(dY^i)^2\,.
\end{align}
The Poincar\'e coordinates on Euclidean $\text{AdS}_{d+1}$ are embedded into $Y\cdot Y=-1$ as
\begin{align}
Y^+=\frac{1}{\eta}\,, \;\;\; Y^-=\frac{\eta^2+x^2}{\eta}\,, \;\;\; Y^i=\frac{x^i}{\eta}\,,
\end{align} 
and Euclidean space $\mathbb{R}^{d}$ on the AdS boundary is embedded into $X\cdot X=0$ as
\begin{align}
X^+=1\,, \;\;\; X^-=x^2\,, \;\;\; X^i=x^i\,,\label{bce}
\end{align}
where we choose a gauge condition $X^+=1$. 

With this convention, the geodesic (\ref{geo}) is lifted to 
\begin{align}
Y(\lambda)=\frac{e^{-\lambda}X_a+e^{\lambda}X_b}{\sqrt{-2X_a\cdot X_b}}\,,\label{ge}
\end{align}
and the AdS scalar propagators are given by
\begin{align}
G_{b\partial}\left(Y, X_a; \Delta_a\right)&=\mathcal{C}_{\Delta_a, 0}\frac{1}{(-2Y\cdot X_a)^{\Delta_a}}\,,\label{efbpp}\\
G_{bb}\left(Y, Y'; \Delta, 0\right)&=\mathcal{C}_{\Delta, 0}\left(\frac{\xi}{2}\right)^\Delta\,_2F_1\left(\frac{\Delta}{2}, \frac{\Delta+1}{2}, \Delta+1-\frac{d}{2}; \xi^2\right)\,,  \label{efbbp} \\
\mathrm{where}\quad \xi&=-\frac{1}{Y\cdot Y'}\,.\label{efbbp11}
\end{align}

In the embedding formalism, we can rewrite the geodesic Witten diagrams $\mathcal W_{\Delta,\ell}$ (\ref{gwd}) as
\begin{equation}
\begin{split}
\mathcal W_{\Delta,\ell}\,=\,\int_{\gamma_{12}} d\lambda \int_{\gamma_{34}} d\lambda' &G_{b\partial}\left(Y(\lambda), X_1; \Delta_1\right)G_{b\partial}\left(Y(\lambda), X_2; \Delta_2\right)\\
&\times G_{bb}\left(Y(\lambda), Y(\lambda'); \frac{d Y(\lambda)}{d\lambda}, \frac{d Y(\lambda')}{d\lambda'}; \Delta, \ell\right)\\
&\quad\times G_{b\partial}\left(Y(\lambda'), X_3; \Delta_3\right)G_{b\partial}\left(Y(\lambda'), X_4; \Delta_4\right)\,,\label{egwdspinl}
\end{split}
\end{equation}
where we define $G_{bb}\left(Y(\lambda), Y(\lambda'); \frac{d Y(\lambda)}{d\lambda}, \frac{d Y(\lambda')}{d\lambda'}; \Delta, \ell\right)$ by the pulled-back bulk-to-bulk propagators of spin-$\ell$ symmetric traceless fields. While we will compute \eqref{egwdspinl} in more detail in the following subsection, here we perform a simpler computation in an approximate way, which will be more useful when we consider a Lorentzian spacetime in Subsection \ref{subsec:spinl}.

It turned out\cite{Costa:2014kfa, Chen:2017yia} that the pulled-back propagators $G_{bb}$ can be constructed from the so-called index-free polynomials $\Pi_{\Delta, \ell}(Y, Y'; W, W')$, where $W$ and $W'$ are polarization vectors. A systematic construction method of $\Pi_{\Delta, \ell}(Y, Y'; W, W')$ is developed in  \cite{Costa:2014kfa}, and their exact expressions are complicated\footnote{See Subsection \ref{subsec:spinl} for the examples with $\ell=1$ and $\ell=2$.}. 
However, we are interested only in the asymptotic limit $(z,\bar{z}\to0)$, and in this case, one can use simpler expressions, i.e., asymptotic forms of $\Pi_{\Delta, \ell}(Y, Y'; W, W')$ as $Y'$ approaches to the AdS boundary. In this limit, $-Y\cdot Y'$ goes to infinity, and $\Pi_{\Delta, \ell}(Y, Y'; W, W')$ behave like the bulk-to-boundary propagators \cite{Dyer:2017zef, Costa:2014kfa}
\begin{equation}
\begin{split}
&\Pi_{\Delta, \ell}(Y, Y'; W, W')\\
\,\simeq\,& \mathcal{C}_{\Delta, \ell}\frac{(2(W\cdot Y')(W'\cdot Y)-2(Y\cdot Y')(W\cdot W'))^{\ell}}{(-2Y\cdot Y')^{\Delta+\ell}} \;\;\;(-Y\cdot Y'\to\infty)\,.\label{afbubupe}
\end{split}
\end{equation}
We note that this approximation is valid even though $\ell=0$.

We want to obtain the asymptotic behaviors of $\Pi_{\Delta, \ell}(Y, Y'; \frac{d Y(\lambda)}{d\lambda}, \frac{d Y(\lambda')}{d\lambda'})$. From (\ref{sp}), (\ref{bce}), and (\ref{ge}) the behavior of the numerator of \eqref{afbubupe} is obtained by
\begin{equation}
\begin{split}
&\left(\frac{d Y(\lambda)}{d\lambda}\cdot Y(\lambda')\right)\left(\frac{d Y(\lambda')}{d\lambda'}\cdot Y(\lambda)\right)-\biggl(Y(\lambda)\cdot Y(\lambda')\biggr)\left(\frac{d Y(\lambda)}{d\lambda}\cdot \frac{d Y(\lambda')}{d\lambda'}\right)\\
\,=\,&\frac{w_1\bar{w}_1(-z-\bar{z}+z\bar{z})+w_1(z-z\bar{z})+\bar{w}_1(\bar{z}-z\bar{z})}{w_1\bar{w}_1z\bar{z}}\,\simeq\,-\frac{z+\bar{z}}{z\bar{z}} \;\;\; (z,\bar{z}\to0)\,,\label{cosp}
\end{split}
\end{equation}
where we set $x_1=(w_1, \bar{w}_1, \mathbf{0}_\perp)$ and take $|w_1|\to\infty$ and $|\bar{w}_1|\to\infty$.
 The behavior of the denominator of \eqref{afbubupe} is obtained by (\ref{giscl}) and (\ref{efbbp11}).
As a result, the asymptotic behaviors of $\Pi_{\Delta, \ell}(Y, Y'; \frac{d Y(\lambda)}{d\lambda}, \frac{d Y(\lambda')}{d\lambda'})$ are given by
\begin{align}
\Pi_{\Delta, \ell}\left(Y, Y'; \frac{d Y(\lambda)}{d\lambda}, \frac{d Y(\lambda')}{d\lambda'}\right)\propto (z\bar{z})^{\frac{1}{2}(\Delta-\ell)}(z+\bar{z})^{\ell} \;\;\; (z, \bar{z}\to 0)\,.\label{abslp}
\end{align}

Note that these behaviors agree with the highest order terms of the conformal blocks (\ref{abec}): 
\begin{equation}
    (z\bar{z})^{\frac{\Delta}{2}}C_{\ell}^{d/2-1}\left(\frac{z+\bar{z}}{2\sqrt{z\bar{z}}}\right) \simeq \frac{\Gamma(d/2+\ell-1)}{\Gamma(d/2-1)\Gamma(\ell+1)}(z\bar{z})^{\frac{1}{2}(\Delta-\ell)}(z+\bar{z})^{\ell} \,,
\end{equation}
so we find that $\Pi_{\Delta, \ell}(Y, Y'; \frac{d Y(\lambda)}{d\lambda}, \frac{d Y(\lambda')}{d\lambda'})$ can capture the asymptotic behaviors of the geodesic Witten diagrams and the corresponding conformal blocks.

However, to compare \eqref{abslp} with \eqref{abec}, we first need to do two things. First, we need to construct $G_{bb}\left(Y(\lambda), Y(\lambda'); \frac{d Y(\lambda)}{d\lambda}, \frac{d Y(\lambda')}{d\lambda'}; \Delta, \ell\right)$ from $\Pi_{\Delta, \ell}(Y,$ $Y'; \frac{d Y(\lambda)}{d\lambda}, \frac{d Y(\lambda')}{d\lambda'})$  by using projectors for symmetric traceless fields\footnote{See a connection between the Gegenbauer polynomials and two-point functions with the projectors in \cite{Kobayashi:2020kgb, Costa:2011dw}.}.
Instead of doing such a complicated computation, as a short cut to our goal, we simply use $\Pi_{\Delta, \ell}(Y, Y'; \frac{d Y(\lambda)}{d\lambda}, \frac{d Y(\lambda')}{d\lambda'})$ because it 
can capture the highest order term in the  Gegenbauer polynomial. 
Second, we have to show that the integrations over $\lambda$ and $\lambda$ are factored out independent of the $z\bar{z}$-dependence in the asymptotic behaviors. It is proven in Subsection \ref{sec22} for $\ell=0$, and the same argument works for non-zero $\ell$.

Let us discuss why (\ref{abslp}) does not capture all terms in the Gegenbauer polynomial. Since $\frac{d Y(\lambda)}{d\lambda}$ does not satisfy the traceless condition $W\cdot W=0$ \cite{Costa:2014kfa}, $\Pi_{\Delta, \ell}\left(Y, Y'; \frac{d Y(\lambda)}{d\lambda}, \frac{d Y(\lambda')}{d\lambda'}\right)$ are not suitable for the propagators of traceless fields and may include contributions from lower spin fields\footnote{For example, a spin-2 symmetric field can be decomposed into a spin-2 symmetric traceless part and a scalar part.}. Due to the propagation of such lower spin fields, $\Pi_{\Delta, \ell}\left(Y, Y'; \frac{d Y(\lambda)}{d\lambda}, \frac{d Y(\lambda')}{d\lambda'}\right)$ without the projection cannot provide all terms in the Gegenbauer polynomial.


\subsection{Asymptotic behaviors of conformal blocks from the classical equations}\label{secec}

In the previous subsection, we obtained the highest order term of the asymptotic behavior of the conformal block (\ref{abec}) by using the index-free polynomials. In this subsection, we obtain (\ref{abec}) more precisely.
Even if we do not know the exact expressions of the bulk-to-bulk propagators, we can derive the asymptotic behaviors (\ref{abec}) by analyzing the classical equations of free fields on the geodesics. This can be done because one of the integrations over $\lambda$ on the left geodesic curve $\gamma_{12}$ in the amplitude of the geodesic Witten diagrams \eqref{gwd} can be written as   
\begin{equation}
\begin{split}
 &\int_{\gamma_{12}} d\lambda\; G_{b\partial}\left(y(\lambda), x_1; \Delta_1\right)G_{b\partial}\left(y(\lambda), x_2; \Delta_2\right)G_{bb}\left(y(\lambda), y(\lambda'); \Delta, \ell\right)\\
 &\quad =h_{\mu_1\dots \mu_\ell}(y(\lambda'))\frac{d y^{\mu_1}(\lambda')}{d\lambda'}\dots\frac{d y^{\mu_\ell}(\lambda')}{d\lambda'}\, , \label{pbslf}
\end{split}
\end{equation} 
which is equal to the pulled-back spin-$\ell$ field   with a normalization that depends on $x_1$, $x_2$, $\Delta_1$, and $\Delta_2$ \cite{Hijano:2015zsa}. Thus, the geodesic Witten diagrams $\mathcal W_{\Delta,\ell}$ \eqref{gwd} become
\begin{equation}
\begin{split}
	\mathcal W_{\Delta,\ell}=\int_{\gamma_{34}}d\lambda'\;&h_{\mu_1\dots \mu_\ell}(y(\lambda'))\frac{d y^{\mu_1}(\lambda')}{d\lambda'}\dots\frac{d y^{\mu_\ell}(\lambda')}{d\lambda'}\\
	&\times G_{b\partial}\left(y(\lambda'), x_3; \Delta_3\right)G_{b\partial}\left(y(\lambda'), x_4; \Delta_4\right)\,.
\end{split}
\end{equation}

By using the equivalence of $W_{\Delta,\ell}$ \eqref{cbwp} and $\mathcal W_{\Delta,\ell}$ \eqref{gwd}, we have
\begin{equation}
	G_{\Delta,\ell}(z,\bar z)\propto \int_{\gamma_{34}}d\lambda'\;h_{\mu_1\dots \mu_\ell}(y(\lambda'))\frac{d y^{\mu_1}(\lambda')}{d\lambda'}\dots\frac{d y^{\mu_\ell}(\lambda')}{d\lambda'}e^{-(\Delta_3-\Delta_4)\lambda'}\, , \label{euclidean conformal block from eom precise}
\end{equation}
where the prefactors in \eqref{cbwp} are canceled by the bulk-to-boundary propagators as we saw in \eqref{B2b result}. It turns out that we do not need to integrate over $\lambda'$ along the geodesic line $\gamma_{34}$ in \eqref{euclidean conformal block from eom precise} to determine the $z\bar{z}$-dependence in the asymptotic behaviors of $G_{\Delta,\ell}(z,\bar z)$. This is because, for the leading order term in $z,\bar z\rightarrow0$ limit, the $\lambda'$-dependent part in $y(\lambda')$ can be separated from the others as shown in \eqref{giscl}. Thus, the $\lambda'$ integration can be integrated out from the leading term in $z,\bar z\to0$ limit and does not affect the asymptotic behaviors in terms of $z$ and $\bar z$. In other words, the $z\bar{z}$-dependence in the asymptotic behaviors of $G_{\Delta,\ell}(z,\bar z)$ can be solely determined from the $z\bar z$-dependence of the field $h_{\mu_1\dots\mu_\ell}$ contracted with $d y^{\mu}(\lambda')/d\lambda'$s:
\begin{equation}
	G_{\Delta,\ell}(z,\bar z)\propto \left( h_{\mu_1\dots \mu_\ell}(y(\lambda'))\frac{d y^{\mu_1}(\lambda')}{d\lambda'}\dots\frac{d y^{\mu_\ell}(\lambda')}{d\lambda'}\right)(z,\bar z)\, . \label{euclidean conformal block from eom}
\end{equation}

First, let us consider the geodesic Witten diagram $\mathcal W_{\Delta,0}$ with the scalar field  in the bulk. To reproduce the asymptotic behavior of the conformal block $G_{\Delta,0}(z, \bar{z})$ by using \eqref{euclidean conformal block from eom} with $\ell=0$, we only need to take care of the asymptotic behavior of scalar field $h(y(\lambda'))$ in $z,\bar z\rightarrow0$ limit. The asymptotic behavior of the free scalar field $h(y(\lambda'))$ in the bulk with mass squared $m^2=\Delta(\Delta-d)$ can be determined by its equation of motion, 
\begin{align}
\left(\nabla_\mu\nabla^\mu-\Delta(\Delta-d)\right)h(y(\lambda'))=0\, ,\label{seom}
\end{align}
where $\nabla_\mu$ is the covariant derivative with respect to $y(\lambda')$. As well known in the study of AdS/CFT, (\ref{seom}) has two asymptotic solutions in small $\eta(\lambda')$:
\begin{align}
h(y(\lambda'))\propto \eta(\lambda')^\Delta \;\;\; \text{or} \;\;\; \eta(\lambda')^{d-\Delta} \;\;\; (\eta(\lambda')\to0)\, .
\end{align}
Thus, the normalizable mode $\eta(\lambda')^\Delta\propto(z\bar{z})^{\frac{1}{2}\Delta}$ corresponds to the asymptotic behavior of conformal block $G_{\Delta,0}(z, \bar{z})$ (\ref{abec}), and the non-normalizable mode $\eta(\lambda')^{d-\Delta}\propto(z\bar{z})^{\frac{1}{2}(d-\Delta)}$ corresponds to the one of shadow conformal block $G_{d-\Delta, 0}(z, \bar{z})$.

Next, we consider the geodesic Witten diagrams $\mathcal W_{\Delta,\ell}$ with the symmetric traceless spin-$\ell$ fields $h_{\mu_1\dots \mu_\ell}$ in the bulk. To get the correct asymptotic behaviors of conformal blocks in \eqref{euclidean conformal block from eom}, we have to consider the asymptotic behaviors of the contraction between $h_{\mu_1\dots \mu_\ell}$ and $\frac{d y^{\mu_1}(\lambda')}{d\lambda'}\dots\frac{d y^{\mu_\ell}(\lambda')}{d\lambda'}$. At first, the asymptotic behaviors of the symmetric traceless spin-$\ell$ field solutions with mass squared $m^2=\Delta(\Delta-d)-\ell$ are determined from the equations  \cite{Costa:2014kfa},
\begin{gather}
\left(\nabla_\mu\nabla^\mu -\Delta(\Delta-d)+\ell\right)h_{\mu_1\dots \mu_\ell}(y(\lambda'))=0\,, \label{leom}\\[6pt]
\nabla^{\mu_1}h_{\mu_1\dots \mu_\ell}(y(\lambda'))=0\label{lorenz}\,.
\end{gather}
For the moment, we distinguish the component $\eta$ and the other indices $i$ associated to $x^i$ and make an ansatz for the asymptotic series expansion of $h_{i_1\dots i_k\eta\dots\eta}(\eta, x)$ as
\begin{equation}
	h_{i_1\dots i_k\eta\dots\eta}(\eta, x)=\eta^{\alpha(k)}\sum_{j=0}^\infty \eta^j h^{(j)}_{i_1\dots i_k\eta\dots\eta}(x)\, . \label{Euclidean asymptotic ansatz}
\end{equation}
By plugging this ansatz \eqref{Euclidean asymptotic ansatz} into \eqref{lorenz}, we have
\begin{equation}
	\alpha(k-1)=\alpha(k)+1\, .
\end{equation}
Thus, the most dominant field components of $h_{i_1\dots i_k\eta\dots\eta}(\eta, x)$ in $\eta\rightarrow0$ limit is $h_{i_1\dots i_\ell}(\eta,x)$.
In terms of the Poincar\'e coordinates,\footnote{We temporarily use the diagonal Poincar\'e coordinates $ds^2=\frac1{\eta^2}\left(d\eta^2+\sum\limits_{i=1}^{d}(dx^i)^2\right)$, where $y^\mu=(\eta, x^i)=(\eta,\frac12(w+\bar w),\frac1{2i}(w-\bar w)  ,\mathbf x_\perp)$ rather than the non-diagonal coordinates \eqref{poincare} for a simpler expression of \eqref{spin-l poincare eom}.} the equation of the field $h_{i_1\dots i_\ell}$ \eqref{leom} with the condition \eqref{lorenz} becomes \cite{Giombi:2009wh}
\begin{equation}
\begin{split}
	&\left(\nabla_\mu\nabla^\mu -\Delta(\Delta-d)+\ell\right)\, h_{i_1\dots i_\ell}(\eta, x)\\
	&\,=\,\bigg(\eta^2(\partial_\eta+(\ell-d+1)\eta^{-1})(\partial_\eta+\ell\eta^{-1})+\eta^2\delta^{ij}\partial_i\partial_j-\Delta(\Delta-d)\bigg)h_{i_1\dots i_\ell}(\eta, x)\\
	&\;\;\;\;\quad -2\ell\eta\partial_{(i_1}h_{i_2 \dots i_\ell)\eta}(\eta, x)+\ell(\ell-1)\delta_{(i_1   i_2}h_{i_3 \dots i_\ell)\eta\eta}(\eta, x)\, .\label{spin-l poincare eom}
\end{split}
\end{equation}
After we insert the ansatz \eqref{Euclidean asymptotic ansatz} into \eqref{spin-l poincare eom}, we have
\begin{equation}
\begin{split}
	&\left(\nabla_\mu\nabla^\mu -\Delta(\Delta-d)+\ell\right)\, h_{i_1\dots i_\ell}(\eta, x)\\
	&\,=\,\left(\alpha(\ell)-\Delta+\ell\right)\left(\alpha(\ell)-d+\Delta+\ell\right)\eta^{\alpha(\ell)}h^{(0)}_{i_1\dots i_\ell}(x)+\sum_{j=1}^\infty\eta^{\alpha(\ell)+j}\dots\, ,
\end{split}
\end{equation}
and its leading order coefficient of $\eta^{\alpha(\ell)}$ vanishes at $\Delta-\ell$ or $d-\Delta-\ell$, which gives the possible dominant contribution of $h_{i_1\dots i_\ell}(\eta,x)$ as
\begin{equation}
	h_{i_1\dots i_\ell}(\eta,x)\,\propto\, \eta^{\Delta-\ell}\quad\mathrm{or}\quad \eta^{d-\Delta-\ell} \;\;\; (\eta\rightarrow0)\, .
\end{equation}  

The tensor structure of $h_{i_1\dots i_\ell}$ is determined by $x_{a}$ in (\ref{sp}), where $x_1$ diverges, $x_2$ is the zero vector, and $x_3\to x_4$ in $z,\bar z\rightarrow0$ limit.
Thus, we can infer that the finite and nonzero tensor structure of $h_{i_1\dots i_\ell}$ in $z,\bar z\rightarrow0$ limit is determined by $x_4$ as
\begin{align}
h_{i_1\dots i_\ell}\,\propto x_{4j_1}\cdots x_{4j_\ell}\Pi_{i_1\dots i_\ell}^{j_1\dots j_\ell} \;\;\; (z,\bar z\rightarrow0)\,,\label{tshi}
\end{align}
where $\Pi_{i_1\dots i_\ell}^{j_1\dots j_\ell}$ is the projector onto the spin-$\ell$ symmetric traceless fields \cite{Costa:2011dw}.
Also,  from \eqref{geo}, the components of $\frac{d x(\lambda')}{d\lambda'}$ are given by
\begin{equation}
\frac{d x(\lambda')}{d\lambda'}=\left(\frac{z}{2\cosh^2 \lambda'}, \frac{\bar z}{2\cosh^2 \lambda'}, \mathbf{0}_\perp\right)\,.\label{cdxdl}
\end{equation} 

In conclusion,  the dominant contribution comes from the contraction of $i_1,\dots,i_\ell$ in \eqref{euclidean conformal block from eom},
\begin{align}
	G_{\Delta,\ell}\,\propto\, & h_{i_1\dots i_\ell}(y(\lambda'))\frac{d x^{i_1}(\lambda')}{d\lambda'}\dots\frac{d x^{i_\ell}(\lambda')}{d\lambda'} \;\;\; (z,\bar z\rightarrow0)\, .
\end{align}
By using the fact $h_{i_1\dots i_\ell}\;\propto\; \eta^{\Delta-\ell}$, (\ref{tshi}), (\ref{cdxdl}), and a formula of the Gegenbauer polynomial \cite{Costa:2011dw} 
\begin{align}
p_{j_1}\dots p_{j_\ell}\Pi_{i_1\dots i_\ell}^{j_1\dots j_\ell}q^{i_1}\dots q^{i_\ell}=\frac{\Gamma(d/2-1)\Gamma(\ell+1)}{2^\ell\Gamma(d/2+\ell-1)}(p^2q^2)^{\ell/2}C_{\ell}^{d/2-1}\left(\frac{p\cdot q}{\sqrt{p^2q^2}}\right)\, ,
\end{align}
we obtain the final result
\begin{align}
G_{\Delta, \ell}(z, \bar{z})\simeq (z\bar{z})^{\frac{\Delta}{2}}C_{\ell}^{d/2-1}\left(\frac{z+\bar{z}}{2\sqrt{z\bar{z}}}\right) \;\;\; (z, \bar{z}\to 0)\, ,
\end{align}
and this result matches with (\ref{abec}).
For the other choice $h_{i_1\dots i_\ell}\;\propto\; \eta^{d-\Delta-\ell}$, it corresponds to the contribution of the shadow conformal block $G_{d-\Delta,\ell}$.

The behaviors of conformal blocks are controlled by the conformal Casimir equations \cite{Dolan:2003hv}. It was pointed out in \cite{Hijano:2015zsa} that the conformal Casimir equations correspond to the equations of the bulk-to-bulk propagators without a delta function source in the geodesic Witten diagrams. Therefore, the asymptotic behaviors of conformal blocks can be determined from the equations of the bulk-to-bulk propagators on the geodesic as our analysis.

\section{Half-geodesic Witten diagrams in the Rindler-AdS black hole}\label{sec3}
In this section, we construct the half-geodesic Witten diagrams with four external scalar fields in the two-sided Rindler-AdS black hole. Our construction is motivated by holographic computations of the OTOCs using shock wave geometries. We show that the half-geodesic Witten diagrams at late times have the same asymptotic behaviors of conformal blocks in the Regge limit, which are related to the late-time behaviors of conformal blocks in the OTOCs.

\subsection{Review of the Regge conformal blocks}
Before constructing the half-geodesic Witten diagrams in the Rindler-AdS black hole, we briefly review the Regge conformal blocks \cite{Cornalba:2006xm, Caron-Huot:2017vep, Perlmutter:2016pkf}. 
The Regge limit of a function $\mathcal{A}(z, \bar{z})$ is defined as follows. First, consider an analytic continuation of $\mathcal{A}(z, \bar{z})$ by taking
\begin{align}
(1-z)\to e^{-2\pi i}(1-z)\,.\label{accr}
\end{align}
After this analytic continuation, take a limit
\begin{align}
z, \bar{z}\to0 \;\;\; \text{while keeping} \; \left|\frac{\bar{z}}{z}\right|<1 \; \text{fixed}\,.\label{reggel}
\end{align}
Following this procedure, one can define the Regge limit of $\mathcal{A}(z, \bar{z})$.

Let us consider the case of $\mathcal{A}(z, \bar{z})=G_{\Delta, \ell}(z, \bar{z})$, which is a solution of the conformal Casimir equation with an eigenvalue $\Delta(\Delta-d)+\ell(\ell+d-2)$. 
This eigenvalue is invariant under $\Delta\leftrightarrow 1-\ell$, and therefore the equation has another solution $G_{1-\ell, 1-\Delta}(z, \bar{z})$ with an asymptotic behavior
\begin{align}
G_{1-\ell, 1-\Delta}(z, \bar{z})\propto z^{1-\frac{1}{2}(\Delta+\ell)}\bar{z}^{\frac{1}{2}(\Delta-\ell)} \;\;\; (|\bar{z}|\ll|z|\ll1)\,. 
\end{align}
By applying the analytic continuation (\ref{accr}) to $G_{\Delta, \ell}(z, \bar{z})$, one obtains a linear combination of $G_{\Delta, \ell}(z, \bar{z})$ and $G_{1-\ell, 1-\Delta}(z, \bar{z})$.
When $\Delta+\ell>1$, $G_{1-\ell, 1-\Delta}(z, \bar{z})$ is the leading term, and the Regge conformal block $G^{\text{Regge}}_{\Delta, \ell}(z, \bar{z})$ is given by $G_{1-\ell, 1-\Delta}(z, \bar{z})$ in the limit (\ref{reggel}) up to normalization.
The explicit form of $G^{\text{Regge}}_{\Delta, \ell}(z, \bar{z})$ is \cite{Cornalba:2006xm}
\begin{align}
G^{\text{Regge}}_{\Delta, \ell}(z, \bar{z})\propto z^{1-\frac{1}{2}(\Delta+\ell)}\bar{z}^{\frac{1}{2}(\Delta-\ell)}\,_2F_1\left(\Delta-1, \frac{d}{2}-1, \Delta+1-\frac{d}{2}; \frac{\bar{z}}{z}\right)\,.\label{reggecb}
\end{align}
We comment on the condition $\Delta+\ell>1$. This condition is related to the convergence of the integral transformation for light-ray operators \cite{Kravchuk:2018htv}. We will see that it is also related to the convergence of integrals in the half-geodesic Witten diagrams.

\subsection{Our configuration of the OTOCs}
We also review a computation method of the OTOCs \cite{Roberts:2014ifa, Perlmutter:2016pkf, Haehl:2019eae} and explain our configuration. Consider a Euclidean correlation function $\langle W(x_1) W(x_2) V(x_3) V(x_4)\rangle$ in $d$ dimensional Euclidean space $\mathbb{R}^{d}$, where $W(x)$ and $V(x)$ are scalar primary operators with conformal dimensions $\Delta_W$ and $\Delta_V$, respectively. To introduce a periodic Euclidean time, we use a conformal map from $\mathbb{R}^{d}$ to $S^1\times \mathbb{H}^{d-1}$ \cite{Casini:2011kv}, where $S^1$ represents the Euclidean time with period $2\pi$\footnote{In the case of $d=2$, one can use a conformal map from $\mathbb{R}^{2}$ to $S^1\times \mathbb{R}^{1}$ with an arbitrary period.\label{f1}}. Now, $\mathbb{H}^{d-1}$ is a $d-1$ dimensional hyperbolic space with a metric $d\mathbf{x}^2=d \chi^2+\sinh^2\chi d\Omega_{d-2}^2$, where $d\Omega_{d-2}^2$ is a metric of $d-2$ dimensional unit sphere. By using this conformal map and an analytic continuation to a Lorentzian time, one can compute a Lorentzian correlation function 
\begin{align}
\langle W(t_1-i\delta_1, \mathbf{x}_1)  V(t_3-i\delta_3, \mathbf{x}_3) W(t_2-i\delta_2, \mathbf{x}_2) V(t_4-i\delta_4, \mathbf{x}_4)\rangle\,,\label{otoc}
\end{align}
where $t_a$ is the Lorentzian time, $\delta_a$ is the Euclidean time, and $\mathbf{x}_a$ are coordinates of $\mathbb{H}^{d-1}$. For the analytic continuation to (\ref{otoc}), we need to choose an ordering of $\delta_a$ as $\delta_1>\delta_3>\delta_2>\delta_4$, which is related to (\ref{accr}).

In this paper, we consider a configuration of the OTOCs as 
\begin{align}
t_1=t_2=t_W\,, \;\;\; t_3=t_4=t_V\,,& \;\;\; \mathbf{x}_1=\mathbf{x}_2=\mathbf{x}_W\,, \;\;\; \mathbf{x}_3=\mathbf{x}_4=\mathbf{x}_V\,,\label{conf}\\
\delta_1\to\delta_3=\pi\,,& \;\;\; \delta_2\to\delta_4=0\,.\label{difet}
\end{align}
In this configuration, the cross ratios (\ref{cr}) are given by \cite{Roberts:2014ifa, Perlmutter:2016pkf, Haehl:2019eae}
\begin{align}
z=\frac{2}{1-\cosh(t_R-\mathbf{d})}\,, \;\;\; \bar{z}=\frac{2}{1-\cosh(t_R+\mathbf{d})}\,,
\end{align}
where $t_R:=t_W-t_V$, $\cosh \mathbf{d}:=\cosh \mathbf{d}(\mathbf{x}_W, \mathbf{x}_V)$, and $\cosh \mathbf{d}(\mathbf{x}, \mathbf{x}')$ is the geodesic distance between $\mathbf{x}$ and $\mathbf{x}'$ in $\mathbb{H}^{d-1}$ \cite{Cohl_2012}. At late times $t_R\gg1$ fixing $\cosh \mathbf{d}$, the cross ratios behave as
\begin{align}
z\to -4e^{-t_R+\mathbf{d}}\,, \;\;\; \bar{z}\to -4e^{-t_R-\mathbf{d}}\,, \;\;\; \bar{z}/z\to e^{-2\mathbf{d}} \;\;\; (t_R\gg1)\,,\label{ltbcr}
\end{align}
which corresponds to (\ref{reggel}). By substituting (\ref{ltbcr}) to (\ref{reggecb}), we obtain
\begin{align}
G^{\text{Regge}}_{\Delta, \ell}(z, \bar{z})\propto e^{(\ell-1)t_R-(\Delta-1)\mathbf{d}}\,_2F_1\left(\Delta-1, \frac{d}{2}-1, \Delta+1-\frac{d}{2}; e^{-2\mathbf{d}}\right)\,.\label{reggecb2}
\end{align}

We restrict $\mathbf{d}$ to the regime $\mathbf{d}>0$, which is equal to $\mathbf{x}_W\ne \mathbf{x}_V$, for $\left|\frac{\bar{z}}{z}\right|<1$ in (\ref{reggel}) and for the convergence of the hypergeometric function in (\ref{reggecb2}). It is known that the $\mathbf{d}$-dependent part  $F(\mathbf{d}):=e^{-(\Delta-1)\mathbf{d}}\,_2F_1\left(\Delta-1, \frac{d}{2}-1, \Delta+1-\frac{d}{2}; e^{-2\mathbf{d}}\right)$ in (\ref{reggecb2}) is a solution of the following equation \cite{Cornalba:2006xm, Cornalba:2007fs}
\begin{align}
\big[\Box_{\mathbb H}- (\Delta-1)(\Delta-d+1)\big]F(\mathbf{\mathbf{d}})=0\,,\label{esphs}
\end{align}
where $\Box_{\mathbb H}$ is the Laplacian operator on $\mathbb{H}^{d-1}$, and we ignore the delta function by using $\mathbf{d}>0$.

\subsection{Embedding formalism of the Rindler-AdS black hole}
\begin{figure}
\centering
     {\includegraphics[width=5cm]{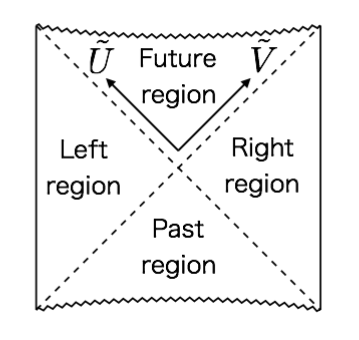}}
 \caption{Penrose diagram of the Rindler-AdS black hole space time with an appropriate transformation $\tilde{U}(U)$ and $\tilde{V}(V)$. The space time is divided into four regions by the horizons at $\tilde{U}=U=0$ and $\tilde{V}=V=0$.  }\label{penrose}
\end{figure}

We summarize the embedding formalism of the Rindler-AdS black hole for systematic computations of the half-geodesic Witten diagrams following \cite{Ahn:2020csv}. In terms of Kruskal coordinates, a metric of the Rindler-AdS black hole is given by
\begin{align}
ds^2=-\frac{4 dU dV}{(1+UV)^2}+\left(\frac{1-UV}{1+UV}\right)^2d\mathbf{x}^2\,.
\end{align}
The Penrose diagram of this Rindler-AdS black hole is shown in Fig.~\ref{penrose}. Schwarzschild coordinates $(t, r)$ of the left and right regions are
\begin{align}
U&=+e^{r_*-t}, \;\;\; V=-e^{r_*+t} \;\;\; (\text{left region})\,,\\
U&=-e^{r_*-t}, \;\;\; V=+e^{r_*+t} \;\;\; (\text{right region})\,,
\end{align}
where tortoise coordinate $r_*$ is $r_*=\log \left(\frac{r-1}{r+1}\right)^{1/2}$. The coordinates of left and right regions are related via $t\leftrightarrow t-i\pi$, which can be  interpreted as the difference of Euclidean time $\pi$ in (\ref{difet}).  In terms of these coordinates, the black hole metric is
\begin{align}
ds^2=-(r^2-1)dt^2+\frac{dr^2}{r^2-1}+r^2d\mathbf{x}^2\,. \label{radsbh metric}
\end{align}
The Hawking temperature of this black hole is $T=1/\beta=1/2\pi$.

This Rindler-AdS black hole\footnote{The BTZ black hole ($d=2$) with any temperature can be also embedded into $\mathbb{R}^{2, 2}$, which is related to the conformal map in Footnote \ref{f1}.} can be embedded into Minkowski space $\mathbb{R}^{2, d}$ with a metric
\begin{align}
ds^2=-(dY^{-1})^2-(dY^{-2})^2+(dY^i)^2\,.\label{lms}
\end{align}
The Kruskal coordinates of the Rindler-AdS black hole are embedded into $Y\cdot Y=-1$ as
\begin{align}
Y^{-1}&=\frac{V+U}{1+UV}\,, \;\;\; Y^{-2}=\frac{1-UV}{1+UV}\cosh \chi\,,\notag\\
  \sum_{i=1}^{d-1}(Y^i)^2&=\left(\frac{1-UV}{1+UV}\right)^2\sinh^2\chi\,, \;\;\; Y^d=\frac{V-U}{1+UV}\,.
\end{align}
Right boundary points at $r=\infty$ are embedded into $X_R\cdot X_R=0$ as
\begin{align}
X_R^{-1}=\sinh t\,, \;\;\; X_R^{-2}=\cosh \chi\,, \;\;\; \sum_{i=1}^{d-1}(X_R^i)^2=\sinh^2\chi\,, \;\;\; X_R^d=\cosh t\,,
\end{align}
and left boundary points at $r=\infty$ are embedded into $X_L\cdot X_L=0$ as
\begin{align}
X_L^{-1}=-\sinh t\,, \;\;\; X_L^{-2}=\cosh \chi, \;\;\; \sum_{i=1}^{d-1}(X_L^i)^2=\sinh^2\chi\,, \;\;\; X_L^d=-\cosh t\,.
\end{align}

We formally use (\ref{efbpp}) and (\ref{efbbp}) with inner products $Y\cdot X$ and $Y\cdot Y'$ on $\mathbb{R}^{2, d}$ (\ref{lms}) as scalar bulk propagators in the Rindler-AdS black hole. Their explicit expressions in terms of the Kruskal and Schwarzschild coordinates are
\begin{align}
G_{b\partial}\left(Y', X_R; \Delta_a\right)&=\frac{\mathcal{C}_{\Delta_a, 0}}{2^{\Delta_a}}\left(\frac{1+U'V'}{+U'e^{t}-V'e^{-t}+(1-U'V')\cosh \mathbf{d}(\mathbf{x}, \mathbf{x}')}\right)^{\Delta_a},\\
G_{b\partial}\left(Y', X_L; \Delta_a\right)&=\frac{\mathcal{C}_{\Delta_a, 0}}{2^{\Delta_a}}\left(\frac{1+U'V'}{-U'e^{t}+V'e^{-t}+(1-U'V')\cosh \mathbf{d}(\mathbf{x}, \mathbf{x}')}\right)^{\Delta_a},\label{bbouprobh}\\
G_{bb}\left(Y, Y'; \Delta, 0\right)&=\mathcal{C}_{\Delta, 0}\left(\frac{\xi}{2}\right)^\Delta\,_2F_1\left(\frac{\Delta}{2}, \frac{\Delta+1}{2}, \Delta+1-\frac{d}{2}; \xi^2\right),\label{bbprorbh}\\
\xi&=\frac{(1+UV)(1+U'V')}{2(UV'+VU')+(1-UV)(1-U'V')\cosh \mathbf{d}(\mathbf{x}, \mathbf{x}')}\,.\label{rbhxi}
\end{align}
In a Lorentzian spacetime, $\xi^2$ can be larger than one. In such a case, the hypergeometric function in the bulk-to-bulk propagator (\ref{bbprorbh}) cannot be defined by the hypergeometric series. To avoid this problem, we introduce two half-geodesics in the left and right regions as the domain of integration. 

\subsection{Scalar exchange half-geodesic Witten diagram}
\begin{figure}
\centering
     {\includegraphics[width=10cm]{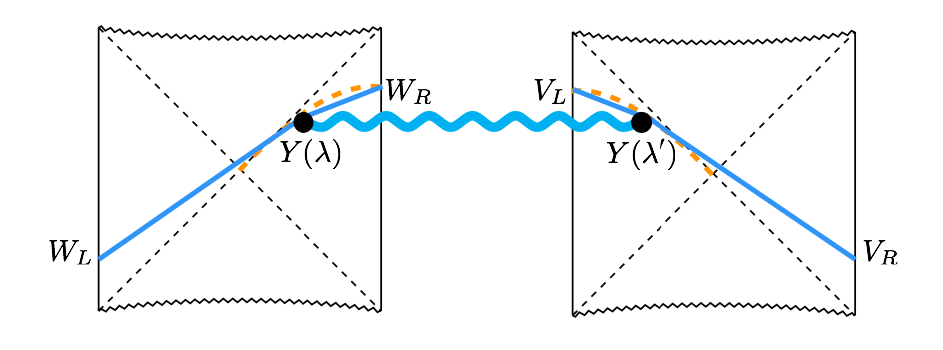}}
 \caption{Half-geodesic Witten diagram in the two-sided Rindler-AdS black hole. The left Penrose diagram is the black hole at $\mathbf{x}=\mathbf{x}_W$, and the right one is at $\mathbf{x}=\mathbf{x}_V$.  Straight lines are the scalar bulk-to-boundary propagators between boundary points $(W_L, W_R, V_L, V_R)$ and bulk points  $\left(Y(\lambda), Y(\lambda')\right)$. A wavy line is a bulk-to-bulk propagator between the bulk points. The bulk points $Y(\lambda)$ and $Y(\lambda')$ are integrated over half-geodesics $(\gamma_W^R, \gamma_V^L)$ that are represented dashed curves.}\label{figuregwdrbh}
\end{figure}

Now, we are ready to construct the half-geodesic Witten diagrams in the two-sided Rindler-AdS black hole $\mathcal W^{\mathcal R}_{\Delta,\ell}$, where $\mathcal R$ stands for the Rindler-AdS black hole metric, with the configuration (\ref{conf}) and (\ref{difet}) by using the embedding formalism. Our purpose is to construct the half-geodesic Witten diagrams that have the Regge behaviors (\ref{reggecb2}). 
We first consider the scalar exchange half-geodesic Witten diagram. In analogy with (\ref{gwd}), its scattering amplitude is defined by
\begin{equation}
\begin{split}
\mathcal W^{\mathcal R}_{\Delta,0}:=\int_{\gamma_W^R} d\lambda \int_{\gamma_V^L} d\lambda' &G_{b\partial}\left(Y(\lambda), W_L; \Delta_W\right)G_{b\partial}\left(Y(\lambda), W_R; \Delta_W\right)\\
&\times G_{bb}\left(Y(\lambda), Y(\lambda'); \Delta, 0\right)\\
&\quad \times G_{b\partial}\left(Y(\lambda'), V_L; \Delta_V\right)G_{b\partial}\left(Y(\lambda'), V_R; \Delta_V\right)\,.\label{gwdrb}
\end{split}
\end{equation} 
Here, $W_L$ and $W_R$ are the left and right boundary points at which $W(t_W, \mathbf{x}_W)$ is inserted as shown in Fig.~\ref{figuregwdrbh}, and $\lambda$ is the proper length parameter of $\gamma_W$, which is the geodesic at $(t_W, \mathbf{x}_W)$ between $W_L$ and $W_R$. Similarly, $V_L$, $V_R$, $\lambda'$,  and $\gamma_V$ are the ones for $V(t_V, \mathbf{x}_V)$. The insertions of $W(t_W, \mathbf{x}_W)$ and $V(t_V, \mathbf{x}_V)$ at the left and right boundaries correspond to (\ref{conf}) and (\ref{difet}). We define (\ref{gwdrb}) for the diagram with $\Delta_1=\Delta_2=\Delta_W$ and $\Delta_3=\Delta_4=\Delta_V$, and one can straightforwardly define the cases with $\Delta_1\ne\Delta_2$ or $\Delta_3\ne\Delta_4$.

\begin{figure}
\centering
     {\includegraphics[width=10cm]{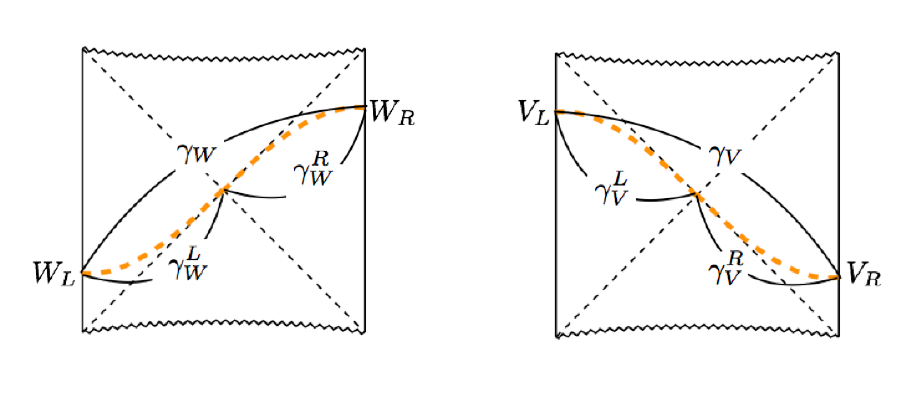}}
 \caption{Geodesics $(\gamma_W, \gamma_V)$ and half-geodesics $(\gamma_W^R, \gamma_W^L, \gamma_V^R, \gamma_V^L)$ are represented schematically as dashed curves.}\label{fig.geodesics}
\end{figure}

Let us explain definitions of $\gamma^R_a$ and $\gamma^L_a$.
Two geodesics $\gamma_W$ and $\gamma_V$ are parametrized as
\begin{align}
\gamma_W:&\;\;\; U(\lambda)=-e^{-t_W}\tanh \frac{\lambda}{2}\,, \;\;\;\; V(\lambda)=e^{t_W}\tanh \frac{\lambda}{2}\,, \;\;\;\; \mathbf{x}(\lambda)=\mathbf{x}_W,\label{gammaw}\\
\gamma_V:&\;\;\; U(\lambda')=-e^{-t_V}\tanh \frac{\lambda'}{2}\,, \;\;\; V(\lambda')=e^{t_V}\tanh \frac{\lambda'}{2}\,, \;\;\; \mathbf{x}(\lambda')=\mathbf{x}_V\,.
\end{align}
See Appendix \ref{appgeo} for details on this parametrization. By using the causal structure of the Penrose diagram, we can naturally divide $\gamma_W$ into two half-geodesics $\gamma_W^R$ and $\gamma_W^L$ between the boundary points and the center of the Penrose diagram.  These half-geodesics are defined as subregions of  $\gamma_W$ such that
\begin{align}
\gamma_W^R=\{\gamma_W | 0\le\lambda<\infty\}\,, \;\;\; \gamma_W^L=\{\gamma_W | -\infty<\lambda\le0\}\,,
\end{align}
and half-geodesics $\gamma_V^R$ and $\gamma_V^L$ are also defined as subregions of  $\gamma_V$ such that
\begin{align}
\gamma_V^R=\{\gamma_V | 0\le\lambda'<\infty\}\,, \;\;\; \gamma_V^L=\{\gamma_V | -\infty<\lambda'\le0\}\,.
\end{align}
See Fig.~\ref{fig.geodesics} for their graphical representations. By using transformations $\lambda\leftrightarrow-\lambda$ and $\lambda'\leftrightarrow-\lambda'$ in (\ref{gwdrb}), one can also define the amplitude of half-geodesic Witten diagram by integration over $\gamma^L_W$ and $\gamma^R_V$.

In the holographic computation of OTOCs, the authors of \cite{Shenker:2013pqa} used the renormalized geodesic distance. Similarly, we need to regularize one of the two integrals in  (\ref{gwdrb}) to obtain the behaviors of conformal blocks. Since $\mathbf{x}$ is held fixed on the geodesics, $\cosh \mathbf{d}(\mathbf{x}(\lambda), \mathbf{x}_W)=\cosh \mathbf{d}(\mathbf{x}(\lambda'), \mathbf{x}_V)=1$ holds in the bulk-to-boundary propagators, and $\cosh \mathbf{d}(\mathbf{x}(\lambda), \mathbf{x}(\lambda'))=\cosh \mathbf{d}$ holds  in the bulk-to-bulk propagator.

We compute the asymptotic behavior of (\ref{gwdrb}) at late times $t_R\gg1$ fixing $\cosh \mathbf{d}$, where $t_R:=t_W-t_V$ and $\cosh \mathbf{d}:=\cosh \mathbf{d}(\mathbf{x}_W, \mathbf{x}_V)$. Without loss of generality, we set $t_W=t_R/2$ and $t_V=-t_R/2$ by using the time translation symmetry. By using the approximation in Appendix \ref{ap:appro} and by introducing new integral variables\footnote{A similar rescaling was used in \cite{Kobayashi:2020kgb}.} 
\begin{align}
\alpha=\frac{2e^{t_R/2}}{\sqrt{2\cosh \mathbf{d}}}\tanh \frac{\lambda}{2}\,, \;\;\; \alpha'=\frac{2e^{t_R/2}}{\sqrt{2\cosh \mathbf{d}}}\tanh  \frac{\lambda'}{2}\,,\label{resiv}
\end{align}
we obtain the asymptotic behavior of $\mathcal W_{\Delta,0}$ (\ref{gwdrb}) in the late-time limit $t_R\gg1$:
\begin{align}
\begin{split}
\mathcal W_{\Delta,0}^{\mathcal R}&\,\simeq\,\frac{\left(\mathcal{C}_{\Delta_W, 0}\mathcal{C}_{\Delta_V, 0}\right)^2\mathcal{C}_{\Delta, 0}}{2^{2(\Delta_W+\Delta_V)+\Delta-1}}\cosh \mathbf{d} e^{-t_R}\\
&\qquad\qquad\times\int_0^\infty d\alpha \int_{-\infty}^0 d\alpha' \xi^\Delta\,_2F_1\left(\frac{\Delta}{2}, \frac{\Delta+1}{2}, \Delta+1-\frac{d}{2}; \xi^2\right)\,\\
&\,=\,\frac{\left(\mathcal{C}_{\Delta_W, 0}\mathcal{C}_{\Delta_V, 0}\right)^2\mathcal{C}_{\Delta, 0}}{2^{2(\Delta_W+\Delta_V)+\Delta-1}} e^{-t_R}\\
&\qquad\qquad\times\int_0^{\infty}d\alpha  \frac{1}{\alpha}\int_0^{1/\cosh \mathbf{d}} d\xi \xi^{\Delta-2}\,_2F_1\left(\frac{\Delta}{2}, \frac{\Delta+1}{2}, \Delta+1-\frac{d}{2}; \xi^2\right),\\
\xi&\,=\,\frac{1}{\cosh \mathbf{d}(1-\alpha\alpha')}\,, \;\;\;d\xi=\xi^2\cosh \mathbf{d}\alpha d\alpha'\,.
\label{abgwdrbh1}
\end{split}
\end{align}
By using formulas of the hypergeometric functions \cite{NIST:DLMF}
\begin{gather}
\begin{split}
&\frac{d}{d\xi}\left(\xi^{\Delta-1}\,_2F_1\left(\frac{\Delta}{2}, \frac{\Delta-1}{2}, \Delta+1-\frac{d}{2}; \xi^2\right)\right)\\
&=(\Delta-1)\xi^{\Delta-2}\,_2F_1\left(\frac{\Delta}{2}, \frac{\Delta+1}{2}, \Delta+1-\frac{d}{2}; \xi^2\right)\,,\notag
\end{split}\\[7pt]
\begin{split}
&(1+e^{-2\mathbf{d}})^{\Delta-1}\,_2F_1\left(\Delta-1, \frac{d}{2}-1, \Delta+1-\frac{d}{2}; e^{-2\mathbf{d}}\right)\\
&=\,_2F_1\left(\frac{\Delta}{2}, \frac{\Delta-1}{2}, \Delta+1-\frac{d}{2}; \frac{4e^{-2\mathbf{d}}}{(1+e^{-2\mathbf{d}})^2}\right)\,,\label{hgff3}
\end{split}\\[7pt]
\begin{split}
\xi^c\,_2F_1\left(\frac{c}{2}, \frac{c+1}{2}, \Delta+1-\frac{d}{2}; \xi^2\right)=&u^{-c}\,_2F_1\left(c, \Delta+\frac{1-d}{2}, 2\Delta+1-d; -\frac{2}{u}\right)\,,\notag \\ &u:=-1+\frac{1}{\xi}\,,
\end{split}
\end{gather}
we obtain
\begin{equation}
\begin{split}
&\int_0^{1/\cosh \mathbf{d}} d\xi \xi^{\Delta-2}\,_2F_1\left(\frac{\Delta}{2}, \frac{\Delta+1}{2}, \Delta+1-\frac{d}{2}; \xi^2\right)\\
=&\frac{1}{\Delta-1}\left(\cosh \mathbf{d}\right)^{1-\Delta}\,_2F_1\left(\frac{\Delta}{2}, \frac{\Delta-1}{2}, \Delta+1-\frac{d}{2}; \left(\cosh \mathbf{d}\right)^{-2}\right)\\
=&\frac{2^{\Delta-1}}{\Delta-1}e^{-(\Delta-1)\mathbf{d}}\,_2F_1\left(\Delta-1, \frac{d}{2}-1, \Delta+1-\frac{d}{2}; e^{-2\mathbf{d}}\right)\,,\label{intxi}
\end{split}
\end{equation}
where we use $\Delta>1$, which is the condition $\Delta+\ell>1$ with $\ell=0$, to derive
\begin{align}
&-\frac{\mathcal{C}_{\Delta, 0}}{2^\Delta(\Delta-1)}\xi^{\Delta-1}\,_2F_1\left(\frac{\Delta}{2}, \frac{\Delta-1}{2}, \Delta+1-\frac{d}{2}; \xi^2\right)\Bigg|_{\xi=0}=H_0(u)\Big|_{u=\infty}=0\,,\label{ccl0}\\
&H_0(u):=-\frac{\mathcal{C}_{\Delta, 0}}{2^\Delta(\Delta-1)}u^{-\Delta+1}\,_2F_1\left(\Delta-1, \Delta+\frac{1-d}{2}, 2\Delta+1-d; -\frac{2}{u}\right)\,,\label{defh}
\end{align}
where we define $H_0(u)$ for later convenience.
Substituting (\ref{intxi}) into (\ref{abgwdrbh1}), we finally obtain the asymptotic behavior of (\ref{gwdrb}) in the late-time limit $t_R\gg1$, 
\begin{equation}
\begin{split}
\mathcal W_{\Delta,0}^{\mathcal R}&\,\simeq\,\frac{\left(\mathcal{C}_{\Delta_W, 0}\mathcal{C}_{\Delta_V, 0}\right)^2\mathcal{C}_{\Delta, 0}}{2^{2(\Delta_W+\Delta_V)}(\Delta-1)} \log \left(\frac{1}{\epsilon}\right)\\
&\qquad\times e^{-t_R-(\Delta-1)\mathbf{d}} \,_2F_1\left(\Delta-1, \frac{d}{2}-1, \Delta+1-\frac{d}{2}; e^{-2\mathbf{d}}\right)\,,\label{abgwdrbh2}
\end{split}
\end{equation}
where we use a regularization 
\begin{align}
\int_{0}^\infty d\alpha \frac{1}{\alpha} \simeq \int_{a\epsilon}^{a} d\alpha \frac{1}{\alpha}=\log \left(\frac{1}{\epsilon}\right)\,, \;\;\; a:=\frac{2e^{t_R/2}}{\sqrt{2\cosh \mathbf{d}}}\,.\label{reg}
\end{align}
The asymptotic behavior (\ref{abgwdrbh2}) agrees with the Regge behavior (\ref{reggecb2}) for the scalar exchange $(\ell=0)$ up to normalization.

\subsection{Spin-$\ell$ exchange half-geodesic Witten diagrams}\label{subsec:spinl}

Just like (\ref{egwdspinl}), scattering amplitude of the spin-$\ell$ exchange half-geodesic Witten diagrams in the two-sided Rindler-AdS black hole $\mathcal W^{\mathcal R}_{\Delta,\ell}$ is defined by
\begin{equation}
\begin{split}
\mathcal W^{\mathcal R}_{\Delta,\ell}:=\int_{\gamma_{W}^R} d\lambda \int_{\gamma_{V}^L} d\lambda' &G_{b\partial}\left(Y(\lambda), W_L; \Delta_W\right)G_{b\partial}\left(Y(\lambda), W_R; \Delta_W\right)\\
&\times G_{bb}\left(Y(\lambda), Y(\lambda'); \frac{d Y(\lambda)}{d\lambda}, \frac{d Y(\lambda')}{d\lambda'}; \Delta, \ell\right)\\
&\qquad\times G_{b\partial}\left(Y(\lambda'), V_L; \Delta_V\right)G_{b\partial}\left(Y(\lambda'), V_R; \Delta_V\right)\,.\label{hgwdspinl}
\end{split}
\end{equation}
From now, we explicitly compute the asymptotic behaviors of (\ref{hgwdspinl}) with $\ell=1$ and $\ell=2$ at late times $t_R\gg1$. In the embedding formalism, the index-free polynomials can be expressed as \cite{Costa:2014kfa}
\begin{align}
&\Pi_{\Delta, \ell}(Y, Y'; W, W')=\sum_{k=0}^\ell(W\cdot W')^{\ell-k}\left((W\cdot Y') (W'\cdot Y)\right)^k g_k^\ell(u)\,,\label{exbbpsl}\\
&g_k^\ell(u)=\sum_{i=k}^\ell(-1)^{i+k}\left(\frac{i!}{k!}\right)^2\frac{1}{(i-k)!}\partial_u^kh_i^\ell(u)\,, \;\;\; u=-1-Y\cdot Y'\,,
\end{align}
where $h_i^\ell(u)$ can be defined by a recursion relation in \cite{Costa:2014kfa}.
The explicit expressions of $h_i^\ell(u)$ with $\ell=1$ and $\ell=2$ are given by
\begin{align}
h_0^{\ell}(u)&=h_0(u):=\mathcal{C}_{\Delta, 0}(2u)^{-\Delta}\;_2F_1\left(\Delta, \Delta+\frac{1-d}{2}, 2\Delta+1-d; -\frac{2}{u}\right)\,,\\
h_1^1(u)&=-\frac{1}{(\Delta-1)(d-\Delta-1)}\left((d-1)h_0(u)+(u+1)\partial_uh_0(u)\right)\,,\\
h_1^2(u)&=-\frac{2}{\Delta(d-\Delta)}\left(dh_0(u)+(u+1)\partial_uh_0(u)\right)\,,\\
h_2^2(u)&=-\frac{1}{2d(\Delta-1)(d-\Delta-1)}\left((d-1)(dh_1^2(u)+(u+1)\partial_uh_1^2(u))+2h_0(u)\right)\,,
\end{align}
where $h_0^{\ell}(u)=h_0(u)$ is the scalar propagator and does not depend on $\ell$.

As discussed in Subsection \ref{subsec:gwdsl}, $\Pi_{\Delta, \ell}(Y, Y'; \frac{d Y(\lambda)}{d\lambda}, \frac{d Y(\lambda')}{d\lambda'})$ without the projection is different from $G_{bb}(Y(\lambda), Y(\lambda'); \frac{d Y(\lambda)}{d\lambda}, \frac{d Y(\lambda')}{d\lambda'}; \Delta, \ell)$ due to lower spin fields. In the Regge limit or the late-time limit, the propagation with the largest spin is dominant as seen in (\ref{reggecb2}). Thus, to compute the asymptotic behaviors when $t_R\gg1$, we use the following approximation
\begin{align}
G_{bb}\left(Y(\lambda), Y(\lambda'); \frac{d Y(\lambda)}{d\lambda}, \frac{d Y(\lambda')}{d\lambda'}; \Delta, \ell\right)\,\simeq\, \Pi_{\Delta, \ell}\left(Y, Y'; \frac{d Y(\lambda)}{d\lambda}, \frac{d Y(\lambda')}{d\lambda'}\right)\,.
\end{align}
We also use the approximation in Appendix \ref{ap:appro} as
\begin{gather}
\begin{split}
\frac{d Y(\lambda)}{d\lambda}\cdot\frac{d Y(\lambda')}{d\lambda'}&\,\simeq\, \frac{e^{t_R}}{2}\,, \\
\left(\frac{d Y(\lambda)}{d\lambda}\cdot Y(\lambda')\right) \left(\frac{d Y(\lambda')}{d\lambda'}\cdot Y(\lambda)\right)&\,\simeq\, \frac{e^{t_R}}{2}\left(\cosh\mathbf{d}-(u+1)\right)\,.
\end{split}\label{approxdYdY}
\end{gather}

First, we compute the asymptotic behavior with $\ell=1$ when $t_R\gg1$. By using the expression of index-free polynomial (\ref{exbbpsl}), \eqref{hgwdspinl} with $\ell=1$ becomes
\begin{equation}
\begin{split}
\mathcal W_{\Delta,1}^{\mathcal R}\,\simeq\,&\frac{\left(\mathcal{C}_{\Delta_W, 0}\mathcal{C}_{\Delta_V, 0}\right)^2}{2^{2(\Delta_W+\Delta_V)}}\\
&\times\int d\alpha \frac{1}{\alpha}\int_{-1+\cosh \mathbf{d}}^\infty du\left(h_0(u)-h_1^1(u)+\left(\cosh\mathbf{d}-(u+1)\right)\partial_uh_1^1(u)\right)\,,\\
1+u&=\cosh \mathbf{d}(1-\alpha\alpha')\,, \;\;\; du=-\cosh \mathbf{d}\alpha d\alpha'\,.
\end{split}
\end{equation}
By using (\ref{hgff3}), (\ref{defh}), and $\partial_uH_0(u)=h_0(u)$, one can evaluate the integration of $h_0(u)$ as in the previous subsection.
After integration by parts, the final result is given by 
\begin{equation}
\begin{split}
\mathcal W_{\Delta,1}^{\mathcal R}\,\simeq\,&\frac{\left(\mathcal{C}_{\Delta_W, 0}\mathcal{C}_{\Delta_V, 0}\right)^2\mathcal{C}_{\Delta, 0}}{2^{2(\Delta_W+\Delta_V)+1}(\Delta-1)}  \log \left(\frac{1}{\epsilon}\right)\\
&\qquad\times e^{-(\Delta-1)\mathbf{d}} \,_2F_1\left(\Delta-1, \frac{d}{2}-1, \Delta+1-\frac{d}{2}; e^{-2\mathbf{d}}\right)\,,\label{abs1}
\end{split}
\end{equation}
where we use $\Delta+1>1$ to derive 
\begin{align}
\left[H_0(u)+\left(\cosh\mathbf{d}-(u+1)\right)h_1^1(u)\right]\Big|_{u=\infty}=0\,.\label{ccl1}
\end{align}
The condition $\Delta+1>1$ for (\ref{ccl1}) with $\ell=1$ is different from the condition $\Delta>1$ for (\ref{ccl0}) with $\ell=0$ due to the additional term proportional to $h_1^1(u)$.

Next, the asymptotic behavior with $\ell=2$ when $t_R\gg1$ is given by
\begin{equation}
\begin{split}
\mathcal W_{\Delta,2}^{\mathcal R}\,\simeq\,&\frac{\left(\mathcal{C}_{\Delta_W, 0}\mathcal{C}_{\Delta_V, 0}\right)^2}{2^{2(\Delta_W+\Delta_V)+1}}e^{t_R} \\
&\times\int d\alpha \frac{1}{\alpha}\int_{-1+\cosh \mathbf{d}}^\infty du\Big(h_0(u)-h_1^2(u)+2h_2^2(u)\\
&\qquad\qquad\qquad\qquad\qquad\quad+\left(\cosh\mathbf{d}-(u+1)\right)(\partial_uh_1^2(u)-4\partial_uh_2^2(u))\\
&\qquad\qquad\qquad\qquad\qquad\quad+\left(\cosh\mathbf{d}-(u+1)\right)^2\partial_u^2h_2^2(u)\Big)\,.
\end{split}
\end{equation}
As well as the case of $\ell=1$, after integrations by parts, we obtain 
\begin{equation}
\begin{split}
\mathcal W_{\Delta,2}^{\mathcal R}\,\simeq\,&\frac{\left(\mathcal{C}_{\Delta_W, 0}\mathcal{C}_{\Delta_V, 0}\right)^2\mathcal{C}_{\Delta, 0}}{2^{2(\Delta_W+\Delta_V)+2}(\Delta-1)}  \log \left(\frac{1}{\epsilon}\right)\\
&\qquad\times e^{t_R-(\Delta-1)\mathbf{d}} \,_2F_1\left(\Delta-1, \frac{d}{2}-1, \Delta+1-\frac{d}{2}; e^{-2\mathbf{d}}\right)\,,\label{abs2}
\end{split}
\end{equation}
where we use $\Delta+2>1$ to derive 
\begin{equation}
\begin{split}
\bigg[H_0(u)+(\cosh\mathbf{d}&-(u+1))(h_1^2(u)-2h_2^2(u))\\
&+\left(\cosh\mathbf{d}-(u+1)\right)^2\partial_u h_2^2(u)\bigg]\Big|_{u=\infty}=0\,.\label{ccl2}
\end{split}
\end{equation}

The asymptotic behaviors (\ref{abs1}) and (\ref{abs2}) agree with the Regge behaviors of conformal blocks (\ref{reggecb2}) with $\ell=1$ and $\ell=2$. In the case of arbitrary spin $\ell$, the index-free polynomials (\ref{exbbpsl}) include $(W\cdot W')^{\ell}h_0(u)$. From the integration of this term, one can obtain the Regge behaviors of conformal blocks. 
More precisely, if the following conditions hold\footnote{See Appendix \ref{sec:ibpsl} for more details.}:
\begin{align}
\left[H_0(u)+\sum_{i=1}^\ell \sum_{j=1}^i\left(\cosh\mathbf{d}-(u+1)\right)^{j}\frac{(-1)^{i+j}i!(i-1)!}{(i-j)!j!(j-1)!}\partial_u^{j-1}h_i^\ell(u)\right]\Bigg|_{u=\infty}=0\,,\label{ccal}
\end{align}
the asymptotic behaviors with the spin-$\ell$ exchange are
\begin{equation}
\begin{split}
\mathcal W_{\Delta,\ell}^{\mathcal R}\,\simeq\,&\frac{\left(\mathcal{C}_{\Delta_W, 0}\mathcal{C}_{\Delta_V, 0}\right)^2\mathcal{C}_{\Delta, 0}}{2^{2(\Delta_W+\Delta_V)+\ell}(\Delta-1)}  \log \left(\frac{1}{\epsilon}\right)\\
&\qquad\times e^{(\ell-1)t_R-(\Delta-1)\mathbf{d}} \,_2F_1\left(\Delta-1, \frac{d}{2}-1, \Delta+1-\frac{d}{2}; e^{-2\mathbf{d}}\right)\,.\label{abas}
\end{split}
\end{equation}
We expect the conditions (\ref{ccal}) to hold when  $\Delta+\ell>1$ and leave a careful analysis of them for future work.

As a consistency check, let us compute the exponential behaviors of (\ref{gwdrb}) and (\ref{hgwdspinl}) in the limit  $t_R\gg\mathbf{d}\gg1$. In the previous computation we  take only $t_R\gg1$ with fixed $\mathbf{d}$. By considering the extra condition for $\mathbf{d}$, the computation becomes simpler. In this limit, one can use the approximation (\ref{afbubupe}) and obtain
\begin{equation}
\begin{split}
&\Pi_{\Delta, \ell}\left(Y, Y'; \frac{d Y(\lambda)}{d\lambda}, \frac{d Y(\lambda')}{d\lambda'}\right)\\
&\,\simeq\,   \mathcal{C}_{\Delta, \ell}\frac{\left(2\left(\frac{d Y(\lambda)}{d\lambda}\cdot Y(\lambda')\right)\left(\frac{d Y(\lambda')}{d\lambda'}\cdot Y(\lambda)\right)-2\Big(Y(\lambda)\cdot Y(\lambda')\Big)\left(\frac{d Y(\lambda)}{d\lambda}\cdot \frac{d Y(\lambda')}{d\lambda'}\right)\right)^{\ell}}{(-2Y(\lambda)\cdot Y(\lambda'))^{\Delta+\ell}}\\
&= \frac{\mathcal{C}_{\Delta, \ell}}{2^{\Delta}}\frac{\left(\cosh t_R \cosh \mathbf{d}\right)^\ell}{(-Y(\lambda)\cdot Y(\lambda'))^{\Delta+\ell}} \,\simeq\, \frac{\mathcal{C}_{\Delta, \ell}}{2^{\Delta+2\ell}}\frac{e^{\ell(t_R+\mathbf{d})}}{(-Y(\lambda)\cdot Y(\lambda'))^{\Delta+\ell}}\,.
\end{split}
\end{equation}
Substituting it into (\ref{hgwdspinl}), we obtain the asymptotic behaviors in the limit  $t_R\gg\mathbf{d}\gg1$ 
\begin{equation}
\begin{split}
\mathcal W_{\Delta,\ell}^{\mathcal R}\,\simeq\,&\frac{\left(\mathcal{C}_{\Delta_W, 0}\mathcal{C}_{\Delta_V, 0}\right)^2\mathcal{C}_{\Delta, \ell}}{2^{2(\Delta_W+\Delta_V)+\Delta+2\ell-1}} e^{(\ell-1)t_R+\ell\mathbf{d}}\int d\alpha \frac{1}{\alpha}\int_0^{2e^{-\mathbf{d}}} d\xi \xi^{\Delta+\ell-2}\\
=&\frac{\left(\mathcal{C}_{\Delta_W, 0}\mathcal{C}_{\Delta_V, 0}\right)^2\mathcal{C}_{\Delta, \ell}}{2^{2(\Delta_W+\Delta_V)+\ell}(\Delta+\ell-1)}\log \left(\frac{1}{\epsilon}\right)e^{(\ell-1)t_R-(\Delta-1)\mathbf{d}}\\
=&\frac{\left(\mathcal{C}_{\Delta_W, 0}\mathcal{C}_{\Delta_V, 0}\right)^2\mathcal{C}_{\Delta, 0}}{2^{2(\Delta_W+\Delta_V)+\ell}(\Delta-1)}\log \left(\frac{1}{\epsilon}\right)e^{(\ell-1)t_R-(\Delta-1)\mathbf{d}}\,,\label{abhgwdspinl}
\end{split}
\end{equation}
where we use $\Delta+\ell>1$, and these behaviors are consistent with (\ref{abgwdrbh2}) and (\ref{abas}). Therefore, we conclude that  (\ref{hgwdspinl}) have the exponential behaviors 
\begin{align}
\mathcal W_{\Delta,\ell}^{\mathcal R}\,\propto\,e^{(\ell-1)t_R-(\Delta-1)\mathbf{d}}\,,\label{ebsl}
\end{align}
in the limit  $t_R\gg\mathbf{d}\gg1$, which agree with (\ref{reggecb2}) in the large spatial distance limit.

\subsection{Similarities with other holographic computations}
Our construction of the half-geodesic Witten diagrams is motivated by holographic computations of the Lyapunov exponent and the butterfly velocity via the shock wave geometry. Their computations in the BTZ and Rindler-AdS black holes \cite{Shenker:2013pqa, Ahn:2019rnq} agree with the exponential behavior of conformal block with the energy-momentum tensor exchange \cite{Roberts:2014ifa, Perlmutter:2016pkf}. We comment on similarities between our computation and the previous computations.

\begin{itemize}
\item At late times $t_W\gg1$, the geodesic $\gamma_W$ approaches the black hole horizon at $U=0$. This reminds us of the shock wave on the horizon created by $W(t_W, \mathbf{x}_W)$. Since the horizon is a null geodesic, our computation in the late-time limit is closely related to the eikonal methods in \cite{Cornalba:2006xk, Cornalba:2006xm, Cornalba:2007zb, Cornalba:2007fs}. Note that our construction of the half-geodesic Witten diagrams is also defined outside the regime $t_W\gg1$.
\item From the viewpoint of half-geodesic Witten diagrams $\mathcal W^{\mathcal R}_{\Delta,\ell}$, the energy-momentum exchange corresponds to the graviton exchange. In particular, (\ref{ebsl}) with the graviton exchange $(\ell=2, \Delta=d)$ agrees with the Lyapunov exponent and the butterfly velocity of the holographic CFTs in the Rindler space \cite{Roberts:2014ifa, Perlmutter:2016pkf, Shenker:2013pqa, Ahn:2019rnq}.
\item In general black holes, one can formally define the half-geodesic Witten diagrams by using propagators for which no analytic expressions are known. Even though we do not know their exact expressions, we may determine the asymptotic behaviors of the half-geodesic Witten diagrams in general black holes by analyzing the asymptotic behaviors of the bulk-to-bulk propagators from the classical equations. In fact, the Lyapunov exponent and the butterfly velocity of theories with Einstein gravity duals in a large class of black holes were studied in \cite{Blake:2016wvh, Roberts:2016wdl}, and similar analysis for the scalar and vector exchange was done in \cite{Kim:2020url}. We note that the butterfly velocity in planar black holes depends on higher derivative couplings \cite{Roberts:2014isa}. This result implies that, in non-maximally symmetric spacetimes, the exponential behaviors cannot be determined from symmetry only, and there is an ambiguity in the choice of propagators and three-point couplings.
\item In our construction, we use the Kruskal coordinates $U$ and $V$. On the other hand, light-cone coordinates in the Poincar\'e patch are used for the construction of the Regge OPE blocks in \cite{Afkhami-Jeddi:2017rmx, Kobayashi:2020kgb}.
\item Our computation is also similar to holographic computations of four-point correlators with two heavy and two light operators, where the two heavy operators make black holes or conical defect geometries in the bulk picture (see, for example, \cite{Fitzpatrick:2014vua, Hijano:2015rla, Fitzpatrick:2015zha,  Hijano:2015qja, Galliani:2016cai, Galliani:2017jlg, Kulaxizi:2018dxo, Giusto:2020mup, Ceplak:2021wak}).
\end{itemize}

\section{Regge behaviors of conformal blocks from the near-horizon analysis}\label{Reggeeomnha}
In Subsection~\ref{secec}, we showed that the asymptotic behaviors (\ref{abec}) can be derived from the equations of the bulk-to-bulk propagators on the geodesic.  One may wonder whether the Regge behaviors could be determined from the equations on the geodesic that approaches the horizon at late times. Here, we derive the equation (\ref{esphs}) for the Regge behaviors from the classical equations and demonstrate that this derivation is related to the near-horizon analysis for the pole-skipping phenomena.

\subsection{
Regge behaviors from the classical equations}\label{Reggeeom}

As shown in Subsection~\ref{secec}, we can determine the asymptotic behaviors of conformal blocks $G_{\Delta,\ell}$ in terms of the field solutions
\begin{equation}
	 G_{\Delta,\ell}\,\propto\,h_{\mu_1\dots \mu_\ell}(y(\lambda))\frac{d y^{\mu_1}(\lambda)}{d\lambda}\dots\frac{d y^{\mu_\ell}(\lambda)}{d\lambda}\, . \label{Lorentzian conformal block from eom}\\[5pt]
\end{equation}
Here, instead of the Euclidean spacetime, we specifically use the Rindler-Ads black hole metric and take the late-time limit $t_R\gg1$ with $t_W=t_R$ and $t_V=0$, which means that  $y(\lambda)$ approaches to the horizon. To describe the bulk field approaching the horizon $r=1$ in a regular way, we use incoming Eddington-Finkelstein coordinates
\begin{equation}
	ds^2\,=\,-(r^2-1)dv^2+2dv dr+r^2d\mathbf x^2\, ,\label{EFmetric}
\end{equation}
where we introduce  
\begin{equation}
v:=r_*+t=\log V\, ,
\end{equation}
to the metric \eqref{radsbh metric}. 

In the embedding formalism, the pulled-back solution \eqref{Lorentzian conformal block from eom} depends on $\frac{d y(\lambda)}{d\lambda}$ through $\frac{d Y(\lambda)}{d\lambda}\cdot\frac{d Y(\lambda')}{d\lambda'}$ and  $\frac{d Y(\lambda)}{d\lambda}\cdot Y(\lambda')$. Also,  $\frac{d}{d\lambda}(V(\lambda)U'(\lambda'))$ in $\frac{d Y(\lambda)}{d\lambda}\cdot\frac{d Y(\lambda')}{d\lambda'}$ and $\frac{d Y(\lambda)}{d\lambda}\cdot Y(\lambda')$ is dominant in the late-time limit $t_R\gg1$ as we saw in \eqref{approxdYdY}. Thus,  the asymptotic behavior of \eqref{Lorentzian conformal block from eom} solely comes  from the dependence of $\frac{d V(\lambda)}{d\lambda}$s as
\begin{equation}
\begin{split}
&h_{\mu_1\dots \mu_\ell}(y(\lambda))\frac{d y^{\mu_1}(\lambda)}{d\lambda}\dots\frac{d y^{\mu_\ell}(\lambda)}{d\lambda}\\
&\,\simeq\, h_{V\dots V}(y(\lambda))\frac{d V(\lambda)}{d\lambda}\dots\frac{d V(\lambda)}{d\lambda}=h_{v\dots v}(y(\lambda))\frac{d v(\lambda)}{d\lambda}\dots\frac{d v(\lambda)}{d\lambda}\, . 
\end{split}
\end{equation}
As $\frac{d v(\lambda')}{d\lambda'}$ does not depend on $t_R$, we can derive the Regge behaviors of conformal blocks from the asymptotic behaviors of the field solutions $h_{v\dots v}$ only.

In terms of the incoming Eddington-Finkelstein coordinates \eqref{EFmetric}, the equations of motion of the symmetric traceless spin-$\ell$ fields $h_{v\dots v}(v,r,\mathbf x)$ become 
\begin{equation}
\begin{split}
	(\nabla_\mu\nabla^\mu&-\Delta(\Delta-d)+\ell) h_{v\dots v}(v,r,\mathbf x)\\
	=&\bigg[r^{-d+1}\partial_r(r^{d-1}(r^2-1)\partial_r)+2 \partial_r\partial_v+(d-1)r^{-1}\partial_v +r^{-2}\Box_{\mathbb H}-\ell d -2\ell r\pa_r\\
	&\quad-\Delta(\Delta-d)+\ell\bigg]h_{v\dots v}(v,r,\mathbf x)+\bigg[2\ell r\partial_v-2\ell(\ell-1)r^2\bigg]h_{v\dots vr}(v,r,\mathbf x)\, ,\label{EFeoml}
\end{split}
\end{equation}
where $\Box_{\mathbb H}$ is the Laplacian operator on the $(d-1)$-dimensional hyperbolic space. To find the field solutions of \eqref{EFeoml} along the horizon in the late-time limit, we make an ansatz of the field solutions that are regular around the horizon $r=1$. Since the Regge behaviors (\ref{reggecb2}) with the spin-$\ell$ exchange are proportional to $e^{(\ell-1)t_R}\propto V^{\ell-1}=e^{(\ell-1)v}$, we make the ansatz\footnote{With this ansatz, it turns out that the coefficient of $h_{v\dots vr}(v,r,\mathbf x)$ term in \eqref{EFeoml} vanishes at the horizon $r=1$.} as  
\begin{equation}
	h_{v\dots v\mu}(v,r,\mathbf x)=e^{(\ell-1) v}\sum_{j=0}^\infty (r-1)^{j} h_{v\dots  v\mu}^{(j)}(\mathbf x)\, .\label{field ansatz}
\end{equation}
By plugging this ansatz into the equations \eqref{EFeoml} and taking $r\rightarrow1$ limit, we have
\begin{equation}
\begin{split}
 (\nabla_\mu\nabla^\mu&-\Delta(\Delta-d)+\ell\left.)h_{v\dots v}(v,r,\mathbf x)\right\vert_{r\rightarrow1}\\
	=&e^{(\ell-1) v}\big[\Box_{\mathbb H}- (d-1)-\Delta(\Delta-d)\big] h^{(0)}_{v\dots v}(\mathbf x)\, .	\label{asympeom}
\end{split}
\end{equation}
Thus, the field solutions satisfy the condition
\begin{equation}
	\left.\big[\Box_{\mathbb H}- (\Delta-1)(\Delta-d+1)\big]h_{v\dots v}(v,r,\mathbf x)\right\vert_{r\rightarrow1}=0\, , \label{Regge from eom}
\end{equation}
and this condition directly gives the equation (\ref{esphs}) for the $\mathbf{d}$-dependence of the Regge behaviors.

In the above computation, we used the ansatz that the solutions are proportional to $V^{\ell-1}$. From the viewpoint of the half-geodesic Witten diagrams, this property comes from the integration of the bulk-to-bulk propagators. To clarify this point, consider an integral of the scalar propagator
\begin{align}
\int_{0}^{1}dU' G(\xi)|_{U=0, V'=0}, \;\;\; G(\xi):=G_{bb}\left(Y, Y'; \Delta, 0\right)\,,
\end{align}
where we evaluate the integral at $U=0$ and $V'=0$ for the late-time behavior. By using (\ref{rbhxi}) and $U'':=\frac{2V}{\cosh \mathbf{d}}U'$, we obtain
\begin{align}
\int_{0}^{1}dU' G(\xi)|_{U=0, V'=0}=\frac{\cosh \mathbf{d}}{2V}\int_0^{\frac{2V}{\cosh \mathbf{d}}}dU''G\left(\frac{1}{\cosh \mathbf{d}(1+U'')}\right)\,.\label{intbbpro}
\end{align}
If the integral converges to a nonzero value at large $V$, (\ref{intbbpro}) is proportional to $V^{-1}$ at large $V$, which corresponds to $V^{\ell-1}$ with $\ell=0$. Since the bulk-to-bulk propagator $G(\xi)$ is a solution of the classical equation\footnote{As long as $\mathbf d>0$, the delta functions in the equations of the bulk-to-bulk propagators can be ignored.}, (\ref{intbbpro}) is also a solution of the equation. We note that the integral measure is important for the $V$-dependence. For example, if we consider an integral with $\int_{0}^{1}dU' U'$, we obtain
\begin{align}
\int_{0}^{1}dU' U' G(\xi)|_{U=0, V'=0}=\left(\frac{\cosh \mathbf{d}}{2V}\right)^2\int_0^{\frac{2V}{\cosh \mathbf{d}}}dU'' U'' G\left(\frac{1}{\cosh \mathbf{d}(1+U'')}\right)\,,
\end{align}
which is proportional to  $V^{-2}$.

\subsection{
Regge behaviors from the near-horizon analysis for the pole-skipping phenomena}

One can also determine the Regge behaviors of conformal blocks by using the near-horizon analysis. The near-horizon analysis detects points so-called pole-skipping points \cite{Blake:2018leo}, where the retarded Green's functions are undetermined at the pole-skipping points. It has been investigated that the retarded Green's function of energy-momentum tensor captures chaotic properties such as the Lyapunov exponent and the butterfly velocity in holographic theories \cite{Grozdanov:2017ajz, Blake:2017ris}.  Also, it has been studied that the pole-skipping points of other spin-$\ell$ fields with $\ell=0,\tfrac12,1,\tfrac32$ can capture similar exponential behaviors (see, for example, \cite{Grozdanov:2019uhi,Blake:2019otz,Natsuume:2019xcy,Ceplak:2019ymw,Yuan:2020fvv,Ceplak:2021efc}).

We formulate the near-horizon analysis of $h_{\mu_1\dots \mu_\ell}$ by using the classical equations \eqref{EFeoml} to determine the Regge behaviors, which is a generalization of the analysis in \cite{Ahn:2020bks}.
For the analysis of the symmetric traceless spin-$\ell$ fields, we make an ansatz 
\begin{equation}
    h_{v\dots v\mu}(v,r,\mathbf x)=e^{-i\omega v}\sum_{j=0}^\infty (r-1)^{j} h_{v\dots v\mu}^{(j)}(\mathbf x)\, .
\end{equation}
By plugging this ansatz into the classical equations \eqref{EFeoml}, at the lowest order $(r-1)^0$, we have
\begin{equation}
\begin{split}
 (\nabla_\mu\nabla^\mu&-\Delta(\Delta-d)+\ell)h_{v\dots v}(v,r,\mathbf x)\\
	=&e^{-i\omega v}\big[\Box_{\mathbb H}-(\ell+i\omega) (d-1)-\Delta(\Delta-d)\big] h^{(0)}_{v\dots v}(\mathbf x)\\
	&-2e^{-i\omega v}\big[i\omega+(\ell-1)\big]h^{(1)}_{v\dots v}(\mathbf x)-2\ell e^{-i\omega v}\big[i\omega+(\ell-1)\big]h^{(0)}_{v\dots vr}(\mathbf x)+\dots\, ,\label{nha lowest}
\end{split}
\end{equation}
where the higher order terms are not shown. 

If we impose the following condition
\begin{equation}
	-i\omega=\ell-1,\qquad \big[\Box_{\mathbb H}- (\Delta-1)(\Delta-d+1)\big]h^{(0)}_{v\dots v}=0\, , \label{Regge from nha}
\end{equation}
 the coefficients of $h^{(0)}_{v\dots v}$ and $h^{(1)}_{v\dots v}$ become zero. Thus, under the condition \eqref{Regge from nha}, $h^{(1)}_{v\dots v}$ cannot be determined in terms of $h^{(0)}_{v\dots v}$ from \eqref{nha lowest}.
Furthermore, the chain of the recurrence relation coming from the higher order terms in \eqref{nha lowest} determines the higher order's coefficients $h^{(j)}_{v\dots v},\ (j>1)$ in terms of the two independent coefficients $h^{(0)}_{v\dots v}$ and $h^{(1)}_{v\dots v}$, which means there are two independent regular solutions.  This is how the near-horizon analysis detects the condition \eqref{Regge from nha}, which determines $e^{(\ell-1)v}\propto e^{(\ell-1)t_R} $ and (\ref{esphs}) for the Regge behaviors (\ref{reggecb2}).

This near-horizon analysis is almost the same as the computation in Subsection \ref{Reggeeom}. However, their methods of determining the $v$-dependence are different. In the near-horizon analysis, we derive $-i\omega=\ell-1$ by imposing that the coefficients of $h^{(0)}_{v\dots vr}$ and $h^{(1)}_{v\dots v}$ are zero. On the other hand, in Subsection \ref{Reggeeom}, the $v$-dependence is determined from the integrals of the bulk-to-bulk propagators in the half-geodesic Witten diagrams $\mathcal W^{\mathcal R}_{\Delta,\ell}$ as in (\ref{intbbpro}). 

\section{Conclusion}\label{sec5}
In this work, we have investigated holographic representations of the Regge conformal blocks by using the Rindler-AdS black hole geometry. We have constructed the half-geodesic Witten diagrams integrated over two half-geodesics with four external scalar fields in the two-sided Rindler-AdS black hole and shown that their late-time behaviors agree with the Regge behaviors of conformal blocks. We have also shown that the near-horizon analysis in the Rindler-AdS black hole, which has been developed in the context of the pole-skipping phenomena, can reproduce the equations for the Regge behaviors.

We summarize several future directions of our work. In our construction, we have considered the specific configuration of the OTOCs as (\ref{conf}) and (\ref{difet}). It would be important to consider the general configurations of the boundary points. In the bulk with Lorentzian signature, the domain of integration for the gravity duals of conformal blocks depends on the causal structure of the boundary points \cite{Kobayashi:2020kgb, Czech:2016xec, deBoer:2016pqk, Chen:2019fvi}. Therefore, the domain of integration for the general configurations may be more complicated than the one in our case. One may construct the amplitude by using retarded or advanced propagators instead of the propagators we have used.

In the original paper of the geodesic Witten diagrams \cite{Hijano:2015zsa}, the authors studied the diagrams with four external scalar fields. Their result has been generalized to the diagrams with various external and exchange fields, for example, \cite{Dyer:2017zef, Chen:2017yia, Nishida:2016vds, Castro:2017hpx, Sleight:2017fpc, Tamaoka:2017jce, Nishida:2018opl, Chen:2020ipe}. One can make such a generalization of our construction as well. In particular, it would be interesting to compare the diagrams with the fermion exchange and the near-horizon analysis of fermion fields \cite{Ceplak:2019ymw, Ceplak:2021efc}.

In this paper, we have only calculated the late-time behaviors of the half geodesic Witten diagrams. By using the projection onto traceless fields, we can evaluate the sub-leading terms in the diagrams.  The near-horizon analysis for the pole-skipping phenomena also can be used to investigate the sub-leading orders \cite{Haehl:2019eae, Grozdanov:2019uhi, Blake:2019otz,  Ahn:2020bks, Ahn:2020baf}. In order to understand more deeply the pole-skipping phenomena,  we should analyze and compare these sub-leading orders. We hope to report on these issues in the near future.

\acknowledgments

We would like to thank Heng-Yu Chen and Viktor Jahnke for
valuable discussions and comments.
This work was supported by Basic Science Research Program through the National Research Foundation of Korea(NRF) funded by the Ministry of Science, ICT \& Future
Planning (NRF- 2021R1A2C1006791) and the GIST Research Institute(GRI) grant funded by the GIST in 2021. 
M. Nishida was supported by Basic Science Research Program through the National Research Foundation of Korea(NRF) funded by the Ministry of Education(NRF-2020R1I1A1A01072726).

\appendix

\section{Geodesics in the two-sided Rindler-AdS black hole}\label{appgeo}
Let us consider a metric 
\begin{align}
ds^2=-f(UV)dUdV\,,\label{gbm}
\end{align}
where we suppress $d\mathbf{x}^2$ because we focus on curves with fixed $\mathbf{x}$. Length of a curve $U=U(V)$ from $(-U_0, -V_0)$ to $(U_0, V_0)$ is given by
\begin{align}
\int_{-V_0}^{V_0}\sqrt{-f(U(V)V)\frac{dU}{dV}}dV\,,
\end{align}
where we assume that $-f(U(V)V)\frac{dU}{dV}$ is not negative. The length is extremized if the following equation of motion holds:
\begin{equation}
\begin{split}
&\frac{d}{dV}\frac{\partial \left(\sqrt{-f(U(V)V)\frac{dU}{dV}}\right)}{\partial \left(dU/dV\right)}-\frac{\partial \left(\sqrt{-f(U(V)V)\frac{dU}{dV}}\right)}{\partial U}\\
=&\frac{1}{2\sqrt{-f \frac{dU}{dV}}}\left(-\frac{1}{2}Uf'+\frac{1}{2}\frac{dU}{dV}Vf'+\frac{1}{2}\frac{f\frac{d^2U}{dV^2}}{\frac{dU}{dV}}\right)=0\,.
\end{split}
\end{equation}
This equation has a solution
\begin{align}
U=\frac{U_0}{V_0}V\,,
\end{align}
and therefore $U=-e^{-2t}V$ is a geodesic in (\ref{gbm}), where we set $U_0/V_0=-e^{-2t}$. In the Rindler-AdS black hole $f(UV)=\frac{4}{(1+UV)^2}$, the geodesic is parametrized by proper length parameter $\lambda$ as
\begin{align}
U(\lambda)=-e^{-t}\tanh \frac{\lambda}{2}\,, \;\;\;\; V(\lambda)=e^{t}\tanh \frac{\lambda}{2}\,.\label{equationgeo}
\end{align}
Substituting (\ref{equationgeo}) into (\ref{gbm}) with $f(UV)=\frac{4}{(1+UV)^2}$, one can check 
\begin{align}
ds^2=d\lambda^2\,.
\end{align}

\section{Approximation in the late-time limit}\label{ap:appro}
In this appendix, we explain the details of the approximation in the late-time limit $t_R\gg1$. By using (\ref{bbouprobh}), (\ref{gammaw}), and (\ref{resiv}), we obtain
\begin{equation}
\begin{split}
G_{b\partial}\left(Y(\lambda), W_L; \Delta_W\right)=&\frac{\mathcal{C}_{\Delta_W, 0}}{2^{\Delta_W}}\left(\frac{1-\left(\tanh \frac{\lambda}{2}\right)^2}{2\tanh \frac{\lambda}{2}+1+\left(\tanh \frac{\lambda}{2}\right)^2}\right)^{\Delta_W}\\
=&\frac{\mathcal{C}_{\Delta_W, 0}}{2^{\Delta_W}}\left(\frac{1-\frac{\cosh \mathbf{d} \alpha^2}{2e^{t_R}}}{\frac{\sqrt{2\cosh \mathbf{d}}\alpha}{e^{t_R/2}}+1+\frac{\cosh \mathbf{d} \alpha^2}{2e^{t_R}}}\right)^{\Delta_W}\,.\label{bbproap}
\end{split}
\end{equation}
In the late-time limit $t_R\gg1$, we use $\alpha/e^{t_R/2}\to0$. With this approximation, (\ref{bbproap}) behaves as
\begin{align}
G_{b\partial}\left(Y(\lambda), W_L; \Delta_W\right)\,\simeq\,\frac{\mathcal{C}_{\Delta_W, 0}}{2^{\Delta_W}}\,.
\end{align}
In the same manner, one can obtain
\begin{equation}
\begin{split}
&\xi=-\frac{1}{Y(\lambda)\cdot Y(\lambda')}\\
&\;\;=\frac{1}{-\frac{1}{2}(e^{t_R}+e^{-t_R})\sinh \lambda \sinh \lambda'+\cosh \mathbf{d} \cosh \lambda \cosh \lambda'}\,\simeq\,\frac{1}{\cosh \mathbf{d}(1-\alpha\alpha')}\,,\\
\end{split}
\end{equation}
\begin{equation}
    1+u=\frac{1}{\xi}\,\simeq\,\cosh \mathbf{d}(1-\alpha\alpha')\,,
\end{equation}
\begin{equation}
    d\alpha=\frac{2e^{t_R/2}}{\sqrt{2\cosh \mathbf{d}}}\frac{1}{2\left(\cosh \frac{\lambda}{2}\right)^2}d\lambda\,\simeq\,\frac{e^{t_R/2}}{\sqrt{2\cosh \mathbf{d}}}d\lambda\,,
\end{equation}
\begin{equation}
    \frac{d Y(\lambda)}{d\lambda}\cdot\frac{d Y(\lambda')}{d\lambda'}=\frac{1}{2}(e^{t_R}+e^{-t_R})\cosh \lambda \cosh \lambda'-\cosh \mathbf{d}\sinh \lambda \sinh \lambda'\,\simeq\,\frac{e^{t_R}}{2}\,,
\end{equation}
\begin{equation}
\begin{split}
&\left(\frac{d Y(\lambda)}{d\lambda}\cdot Y(\lambda')\right) \left(\frac{d Y(\lambda')}{d\lambda'}\cdot Y(\lambda)\right)\\
&=\left(\frac{1}{2}(e^{t_R}+e^{-t_R})\cosh \lambda \sinh \lambda'-\cosh \mathbf{d}\sinh \lambda \cosh \lambda'\right)\\
&\quad\times\left(\frac{1}{2}(e^{t_R}+e^{-t_R})\sinh \lambda \cosh \lambda'-\cosh \mathbf{d}\cosh \lambda \sinh \lambda'\right)\\
&\simeq\,\frac{e^{t_R}}{2}\cosh \mathbf{d}\alpha\alpha'\,\simeq\,\frac{e^{t_R}}{2}(\cosh \mathbf{d}-(u+1))\,.
\end{split}
\end{equation}

\section{Integrals for the spin-$\ell$ exchange}\label{sec:ibpsl}
The late-time behaviors of (\ref{hgwdspinl}) include the following integrals
\begin{equation}
\begin{split}
&\int_{-1+\cosh \mathbf{d}}^\infty du \sum_{k=0}^\ell \left(\cosh\mathbf{d}-(u+1)\right)^k g_k^\ell(u)\\
=&\int_{-1+\cosh \mathbf{d}}^\infty du \sum_{k=0}^\ell \sum_{i=k}^\ell\left(\cosh\mathbf{d}-(u+1)\right)^k (-1)^{i+k}\left(\frac{i!}{k!}\right)^2\frac{1}{(i-k)!}\partial_u^kh_i^\ell(u)\\
=&\int_{-1+\cosh \mathbf{d}}^\infty du \sum_{i=0}^\ell \sum_{k=0}^i\left(\cosh\mathbf{d}-(u+1)\right)^k (-1)^{i+k}\left(\frac{i!}{k!}\right)^2\frac{1}{(i-k)!}\partial_u^kh_i^\ell(u)\,.
\end{split}
\end{equation}
By using integrations by parts, we obtain
\begin{equation}
\begin{split}
&\int_{-1+\cosh \mathbf{d}}^\infty du \sum_{k=0}^\ell \left(\cosh\mathbf{d}-(u+1)\right)^k g_k^\ell(u)\\
=&\int_{-1+\cosh \mathbf{d}}^\infty du \sum_{i=0}^\ell \sum_{k=0}^i \frac{(-1)^{i+k}\left(i!\right)^2}{k!(i-k)!}h_i^\ell(u)\\
&+\left[\sum_{i=1}^\ell \sum_{k=1}^i\sum_{j=1}^k\left(\cosh\mathbf{d}-(u+1)\right)^{k-j+1}\frac{(-1)^{i+k}\left(i!\right)^2}{k!(k-j+1)!(i-k)!}\partial_u^{k-j}h_i^\ell(u)\right]\Bigg|_{u=\infty}\\
=&\int_{-1+\cosh \mathbf{d}}^\infty du h_0(u)
\\
&+\left[\sum_{i=1}^\ell \sum_{j=1}^i\sum_{k=j}^i\left(\cosh\mathbf{d}-(u+1)\right)^{j}\frac{(-1)^{i+k}\left(i!\right)^2}{k!j!(i-k)!}\partial_u^{j-1}h_i^\ell(u)\right]\Bigg|_{u=\infty}\\
=&-H_0(-1+\cosh \mathbf{d})\\
&+\left[H_0(u)+\sum_{i=1}^\ell \sum_{j=1}^i\left(\cosh\mathbf{d}-(u+1)\right)^{j}\frac{(-1)^{i+j}i!(i-1)!}{(i-j)!j!(j-1)!}\partial_u^{j-1}h_i^\ell(u)\right]\Bigg|_{u=\infty}\,,
\end{split}
\end{equation}
where we use
\begin{align}
\sum_{k=j}^i\frac{(-1)^k}{k!(i-k)!}=\frac{(-1)^jj}{i (i-j)!j!} \;\;\; (i\ne0)\,.
\end{align}


\bibliography{HyunSikRefs}
\bibliographystyle{JHEP}

\end{document}